\documentclass[aps,prb,twocolumn,citeautoscript,superscriptaddress,showpacs,english]{revtex4-1}

\usepackage{amsmath}
\usepackage{amssymb}
\usepackage{graphicx}
\usepackage{xcolor}
\usepackage{upgreek}
\usepackage{url}
\usepackage[utf8]{inputenc}
\usepackage{multirow}
\usepackage{tabularx}
\usepackage[para,online,flushleft]{threeparttable}
\usepackage{booktabs}    
\usepackage{dcolumn}
\usepackage{float,graphicx}
\usepackage{subcaption}

\newcommand{\ICL}{Department of Materials, Imperial College London, London SW7 2AZ, United Kingdom}
\newcommand{\Bristol}{Centre for Computational Chemistry, School of Chemistry, University of Bristol, Bristol BS8 1TS, United Kingdom}
\newcommand{\UiT}{Hylleraas Centre for Quantum Molecular Sciences, Department of Chemistry, UiT The Arctic University of Norway, N-9037 Troms\o{}, Norway}

\begin{document}

\title{A High Throughput Virtual Screening Approach for Identifying Thermally Activated Delayed Fluorescence-Based Emitters}
\author{Kritam Thapa}  
\email{These authors contributed equally to this work}
\affiliation{\ICL} 
\author{Jennifer I.\ Jones}     
\email{These authors contributed equally to this work}
\affiliation{\Bristol} 
\author{Laura E.\ Ratcliff} 
\email{laura.ratcliff@bristol.ac.uk}
\affiliation{\Bristol}
\affiliation{\UiT}

\date{\today}

 \begin{abstract}
    
Thermally activated delayed fluorescence (TADF) offers the promise of highly efficient organic light emitting diodes (OLEDs), without the heavy metals requirement of the previous generation of OLEDs. However, the design of new TADF emitters is complicated by competing requirements, which require opposing design strategies. High throughput virtual screening (HTVS) approaches, however, offer the possibility of identifying new TADF emitters without necessarily relying on existing design rules. In this work the STONED algorithm [A.~Nigam \emph{et al.}, \emph{Chem.\ Sci.}, 2021, \textbf{12}, 7079] is used to impose random structural mutations starting from a set of twenty parent molecules, composed of both traditional donor-acceptor and multiresonant TADF emitters. Following this, successive filters are applied based on features of the atomic structure through to time-dependent density functional theory calculations. Although the randomised approach proves to be ill-suited to rediscovering existing TADF emitters, the resulting workflow leads to the identification of a number of molecules with promising properties for TADF, across a range of emission colours. \end{abstract}

\maketitle

\section{Introduction}

Organic light emitting diodes (OLEDs) are an efficient and sustainable offering to the display and lighting markets, offering advantages over previous technologies like liquid crystal displays and traditional LEDs. OLEDs function via electroluminescence, meaning they do not require a backlight, making them energy efficient, thin and light, as well as allowing for ``true black". Furthermore, they have improved resolution and colour contrast, wider viewing angles, and faster response times~\cite{Hong2021,Nayak2023}, while also offering the potential for flexible and transparent displays~\cite{Hong2021,Eng2020}. 
In an OLED, electrons (holes) are injected at the cathode (anode) and recombine to form excitons in the emissive layer, subsequently leading to the emission of a photon as the excited molecules return to their ground states \cite{Hong2021,Sun2015,GomezBombarelli2016}. Due to spin statistics, singlet and triplet excitons form in a 1:3 ratio. The first generation of OLED emitters, namely fluorescent emitters, exploit only singlet excitons for light emission, limiting the internal quantum efficiency (IQE) to 25\%. To overcome this limitation, second-generation phosphorescent emitters arose, which are able to harvest both singlet and triplet excitons for light emission, leading to a theoretical IQE of 100\%. This is enabled by strong spin-orbit coupling due to the incorporation of heavy metal elements, making the triplet state emissive~\cite{Gao2022,GomezBombarelli2016}. However, such materials are costly, potentially toxic, and unstable, particularly in the case of blue emitters~\cite{Dias2017}. The latest generation of emitters can also harvest all excitons radiatively, but are fully organic. These involve reverse intersystem crossing (RISC) from the first triplet (T$_{1}$) to the first singlet (S$_{1}$) excited state, a process activated by the input of thermal energy when the energy gap between S$_{1}$ and T$_{1}$, known as $\Delta E_{\mathrm{ST}}$, is sufficiently small, generally on the order of $0.2$~eV or smaller~\cite{Penfold2018}. This results in both ``prompt" and ``delayed" fluorescence, and is hence known as thermally activated delayed fluorescence (TADF)~\cite{Uoyama2012,GomezBombarelli2016}. 

Small $\Delta E_{\mathrm{ST}}$ in TADF emitters is typically achieved \emph{via} spatial separation of the highest occupied molecular orbital (HOMO) and lowest unoccupied molecular orbital (LUMO)~\cite{Sun2015}. The most common approach involves twisted combinations of donor (D) and acceptor (A) constituents in which the HOMO (LUMO) is strongly localised on the D (A). However, increased HOMO-LUMO separation also reduces the oscillator strength of the S$_{1}$ to S$_{0}$ transition ($f_{\mathrm{S}_{1}}$), thereby reducing the electroluminescent efficiency~\cite{Kim2022}, leading to contradicting requirements of both small $\Delta E_{\mathrm{ST}}$ and strong $f_{\mathrm{S}_{1}}$~\cite{Nakanotani2021}. Furthermore, D-A emitters have large excited state structural relaxations, and in turn a broad Stokes shift and reduced colour purity~\cite{Kim2022}. In 2016 a new design strategy arose based on the multiple resonance (MR) effect~\cite{Hatakeyama2016}. Here the frontier orbitals are instead localised on electron-donating and withdrawing atoms, as well as at their ortho and para positions in a fused polycyclic aromatic structure~\cite{Kim2022}. D-A and MR type emitters can be regarded as the two main classes of TADF emitters, though alternative design mechanisms also exist~\cite{Zhao2021}. 
TADF emitter design is therefore complex, and while the efficiencies of TADF OLED emitters have begun to compete with those of phosphorescent emitters~\cite{Olivier2018}, there is still a need for further developments in order to achieve high efficiency, stability and colour purity simultaneously. 

One approach which is gaining traction for identifying new TADF emitters is high-throughput virtual screening (HTVS), whereby candidates are filtered out at successive stages based on computational chemistry-based predictions of properties like $\Delta E_{\mathrm{ST}}$ and $f_{\mathrm{S}_{1}}$.
An example is the work of G\'{o}mez-Bombarelli \textit{et al.}~\cite{GomezBombarelli2016}, who used machine learning (ML) and time-dependent density functional theory (TDDFT)~\cite{Runge1984} to investigate a library of 1.6~million molecules, resulting in a number of experimentally-verified OLED devices. Other examples include the work of Lin \textit{et al.}~\cite{Lin2021}, who performed screening based on ionisation energies and electron affinities, which relate to trap-free electron transport in single-layer OLED devices, a recent HTVS study which focused on high-efficiency hosts for blue OLEDs~\cite{An2025}, and the work of Zhao~\textit{et al.}~\cite{Zhao2021}, who screened a database of known molecular materials for TADF properties. There are also many examples of using ML for property prediction based on structural features~\cite{Tu2022,Shu2015,Bu2023,Tan2022,Ge2024}, which can further reduce the cost of HTVS workflows. 

In this work, we introduce an HTVS approach for identifying potential TADF emitters, beginning with a library of 40,000 molecules generated \emph{via} random structural alterations made to 20 existing ``parent'' TADF emitters using the STONED algorithm~\cite{Nigam2021}. We first present the library generation and screening process, as well as benchmarking calculations for calibrating screening parameters. We then present the workflow results, also highlighting molecules with interesting properties. We finish with a summary and suggestions for expanding this work in future.

\section{Methods}

\begin{figure}[ht]
    \centering
    \includegraphics[scale=0.5]{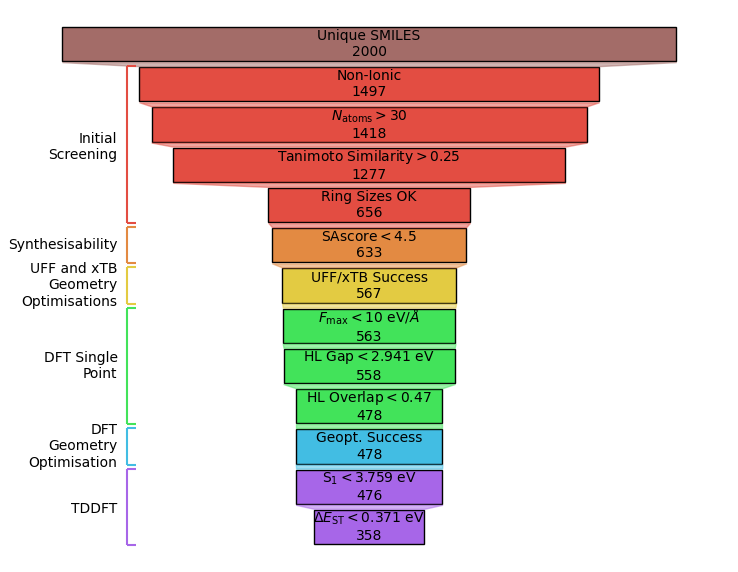}
    \caption{Funnel depiction of the various calculation steps, screening stages and molecules retained at each point for the parent molecule TXO-TPA. Calculation stages are highlighted in a different colour, with individual filters, threshold values and number of molecules passing each stage given on the funnel. For further details see the main text.\label{fig:funnel}}
\end{figure}

A generative approach is used to create a library of ``child'' molecules starting from a known TADF ``parent'' molecule. As is typical of HTVS workflows, a funnel approach is then employed, with molecules screened out at each stage based on filters.  Following library generation, the calculation stages can be broadly separated into initial screening, synthesisability screening, initial geometry optimisations, DFT~\cite{Hohenberg1964,Kohn1965} single point calculations, DFT geometry optimisations and TDDFT calculations. The stages and filters are depicted in Fig.~\ref{fig:funnel}, using the parent molecule TXO-TPA as an example, and are explained in detail in the following. 

\subsection{Library Generation and Initial Filters}

Some previous HTVS studies screen a (subset of a) database of known compounds, the advantage of which being that these are known to be both synthesisable and stable~\cite{Omar2021}. However, TADF emitters may not lie within the chemical space of such databases~\cite{Ge2024}. Therefore, to achieve chemical diversity and increase the likelihood of discovering novel molecules, it can be advantageous to generate a custom library. Previous HTVS studies have done so by combining different D, A and linker moieties~\cite{GomezBombarelli2016,Lin2021,Tu2022,Shu2015,Bu2023,Tan2022,An2025}, though this fails to account for MR structures. Further generations of molecules may be generated from an initial library \emph{via} structural mutations or ML in order to improve diversity~\cite{Shu2015,Tan2022,Rupakheti2016,Ge2024,An2025}. Domain knowledge may also be incorporated into the molecular design process, e.g.\ \emph{via} morphing operations~\cite{Ge2024}. This can hold advantages over randomised approaches in the sense of more relevant molecules being generated, but may lack the novelty of random generation. 

In this work a random approach is taken based on mutating known TADF emitters, balancing the advantages of randomness and domain knowledge. This is done using the STONED algorithm~\cite{Nigam2021}, which produces child molecules through random point mutations to a parent molecule. STONED first reorders an input SMILES string representing the parent, to improve the diversity of generated structures. The resulting SMILES strings are then converted into SELFIES format, an alternative and robust molecular string representation in which all strings represent valid molecules~\cite{Krenn2022}. In this work a single point mutation is applied, defined as a string substitution corresponding to the addition, deletion or replacement of a single SELFIES character. The mutated SELFIES strings are then converted back into SMILES format and canonicalised. For each parent molecule used in this research, 2000 unique child SMILES strings were generated. 

Following molecular generation, a number of rudimentary filters are imposed to refine the search space. First, open-shell molecules are removed, as the focus is on traditional uncharged TADF emitters, and they can also require alternative methods at subsequent workflow steps. Next, a ring size filter is imposed, whereupon candidates without rings or containing rings of sizes outside of 5- or 6- members are removed, since smaller rings are subject to increased ring strain which can negatively affect synthetic accessibility~\cite{Sauve2022} and lead to reduced stability. Both smaller (3-4 member) and larger (7+ member) rings are also scarcely observed in the literature, while to the best of our knowledge there are no known TADF structures without rings. 

Candidates are then filtered based on molecular size, with a minimum threshold of 30 atoms (including H atoms), in order to remove small fragment-like child molecules. No upper limit is applied since child molecules are not significantly bigger than their parents. 
A similarity filter is also included, using the Tanimoto coefficient, which is a measure of the common structural features between two molecules~\cite{Godden2000}. 
ECFP4 fingerprints~\cite{Rogers2010} are used for the calculation of Tanimoto scores, and both fingerprints and similarity scores are obtained using the RDKit package~\cite{rdkit}. The minimum similarity threshold is set to 0.25, indicating that the child structure shares at least 25\% of structural features with its parent. This ensures that child molecules maintain at least part of the parent molecule structure, and are thus not totally random. 

\subsection{Synthesisability and 3D Structure Generation}

An important yet sometimes overlooked consideration in computational molecular design is synthetic accessibility. Approaches for estimating this may broadly be divided into complexity-based methods which assess the complexity of chemical features, and retrosynthetic methods which take into account the full synthetic process including intermediates~\cite{Skoraczynski2023}. 
Full synthetic pathway planning is time-intensive and thus not suitable for an HTVS approach. ML approaches also exist but again require time and expense for training. Cheminformatics tools are therefore the most appropriate choice for a high-throughput workflow; examples include SCScore~\cite{Skoraczynski2023,Coley2018}, RAScore~\cite{Thakkar2021} and SYBA~\cite{Vorsilak2020}. This work employs SAscore, as implemented in RDKit, which takes into account both complex structural features in the form of a complexity penalty, as well as the frequency of molecular fragments among structures in the PubChem database, representing a balance between the aforementioned classes of synthesisability scoring~\cite{Skoraczynski2023}. Although designed for drug-like molecules~\cite{Ertl2009}, SAscore has been used in previous HTVS approaches~\cite{Tan2022,GomezBombarelli2016}, one of which included experimental validation of candidates~\cite{GomezBombarelli2016}. SAscore ranks molecules on a scale from 1 to 10, with lower scores representing increased ease of synthesis and \emph{vice versa}. In this workflow, a threshold of 4.5 was used, at which SAscore has shown equal or better performance to other cheminformatics-based metrics~\cite{Skoraczynski2023}. 

Following SAscore screening, the SMILES strings are converted to 3D structures using RDKit. The universal force field (UFF)~\cite{Rappe1992}, as implemented in RDKit, is then used to optimise up to 10 conformers, following which the lowest energy structure is retained. Other force fields like the Merck Molecular Force Field (MMFF94)~\cite{Halgren1996a,Halgren1996b,Halgren1996c,Halgren1996d,Halgren1996e} are better suited to organic molecules~\cite{LewisAtwell2021}, but MMFF94 does not support all elements, including B, which is frequently seen in MR emitters. Therefore UFF is used for the initial optimisation, but given its limitations, a further geometry optimisation of the lowest energy UFF-optimised structure is then performed using the GFN2-xTB extended semiempirical tight-binding model~\cite{Bannwarth2019}, which is more expensive than UFF but applicable to a wide range of systems, including organic molecules, and improved the stability of the workflow compared to using UFF alone. A small number of UFF optimisations failed to converge within 10,000 iterations, or led to obviously poor structures, e.g.\ with overlapping atoms or separated into distinct structures; such molecules are eliminated. A handful of molecules are also eliminated due to failed xTB optimisations.

\subsection{DFT and TDDFT Calculations}

Having generated a library of 3D molecular structures, excited state calculations can be performed, but given the relatively high cost of TDDFT, particularly for large molecules, it is desirable to have an intermediate screening stage. Based on observed correlations from Penfold for a selection of TADF emitters~\cite{Penfold2015}, the HOMO-LUMO gap and overlap are therefore used to screen out molecules based on predicted values for $\Delta E_{\mathrm{ST}}$ and S$_1$ respectively,
and so the next workflow stage is single point DFT calculations, which are much cheaper than TDDFT. The corresponding thresholds are based on benchmark calculations described in Section~\ref{sec:benchmarks}.

Following the single point calculations, DFT geometry optimisations are performed to improve the xTB-optimised structures, followed by TDDFT calculations. Despite its limitations, particularly with respect to long range CT excitations, TDDFT is widely used in HTVS workflows due to its balance between accuracy and computational cost. While ultimately adiabatic excitation energies are required for direct comparison with experiment, excited state geometry optimisations are expensive, and thus only vertical excitations are calculated, in line with other workflows, which either focused on vertical excitations only or performed excited state geometry optimisations for select molecules only at a later step~\cite{Shu2015,GomezBombarelli2016,Zhao2021}. As shown in Section~\ref{sec:benchmarks}, this proves to be a reasonable approximation.
Explicit filters are imposed on S$_1$ and $\Delta E_{\mathrm{ST}}$, with the thresholds chosen based on benchmark calculations described below.
No explicit filter is used for $f_{S_1}$, this is instead considered qualitatively for molecules which pass all filters.

\subsubsection{Computational Details.}

Ground state DFT calculations were performed using the BigDFT code~\cite{Ratcliff2020}, using HGH-GTH pseudopotentials\cite{Goedecker1996,Hartwigsen1998}, with non-linear core corrections where available~\cite{Willand2013}. Single point calculations employed a wavelet grid spacing of 0.45~bohr and coarse (fine) radius multipliers of 5 (7). A small number of molecules were screened out due to not converging within 80 self-consistent iterations, as defined by a wavefunction gradient of less than $10^{-4}$. Molecules were also eliminated if the norm of the DFT forces of the xTB-optimised structure was above 10~eV/\AA, indicating a possible convergence problem or a poor structure.  Geometry optimisations employed the same radius multipliers but with a smaller grid spacing of 0.4~bohr, to allow for more accurate force calculations, and employed the semi-empirical dispersion correction scheme of Grimme (DFT-D3)~\cite{Grimme2010}. Molecules were again screened out if after 500 geometry steps the force norm was not below the target threshold of 0.03~eV/\AA. All BigDFT calculations employed the PBE exchange-correlation functional~\cite{Perdew1996}.

TDDFT calculations were performed with the NWChem code~\cite{Valiev2010} in gas phase, using the Tamm-Dancoff approximation (TDA)~\cite{Hirata1999}, the B3LYP exchange-correlation functional~\cite{Becke1993,Vosko1980,Lee1988,Stephens1994} and 6-31G* basis set for all atoms except I, where 6-311G* was used instead. Initial benchmarks were performed for a subset of parent molecules using cc-pVDZ and cc-pVTZ, where it was found that there was only a small difference in energies at the cost of more expensive calculations. Similarly, initial benchmarks performed using the PBE0 functional~\cite{Adamo1999} showed little difference with respect to B3LYP. Although not optimal for either D-A emitters, where e.g.\ optimally tuned range-separated hybrid functionals~\cite{Penfold2015,Sun2015} have been shown to give better agreement with experiment than pure hybrid functionals, or for MR emitters, where TDDFT significantly overestimates $\Delta E_{\mathrm{ST}}$, due to the lack of doubly excited determinants~\cite{Hall2022,Kunze2024}, cost was a key consideration. Furthermore, although e.g.\ absolute $\Delta E_{\mathrm{ST}}$ values are systematically overestimated for MR emitters, trends between molecules are much less severely affected by this limitation, which is sufficient for this work given the aim is to identify candidate child molecules which show promising properties relative to a specific parent molecule, rather than accurately predict $\Delta E_{\mathrm{ST}}$ values. A handful of molecules used for benchmarking and one molecule in the final workflow were screened out after TDDFT due to having negative S$_1$ or T$_1$ energies, which was assumed to have been caused by convergence issues.

All workflow steps were implemented in the form of Jupyter notebooks, where the DFT and TDDFT stages employed the remotemanager~\cite{Dawson2024} Python package to submit calculations and retrieve data from a remote supercomputer. PyBigDFT~\cite{Dawson2024} was used to generate input files and post-process both BigDFT and NWChem results, as well as to wrap xTB calculations.  

\section{Benchmark Calculations}\label{sec:benchmarks}

\begin{figure*}[ht]
    \centering
    \includegraphics[width=1.0\textwidth]{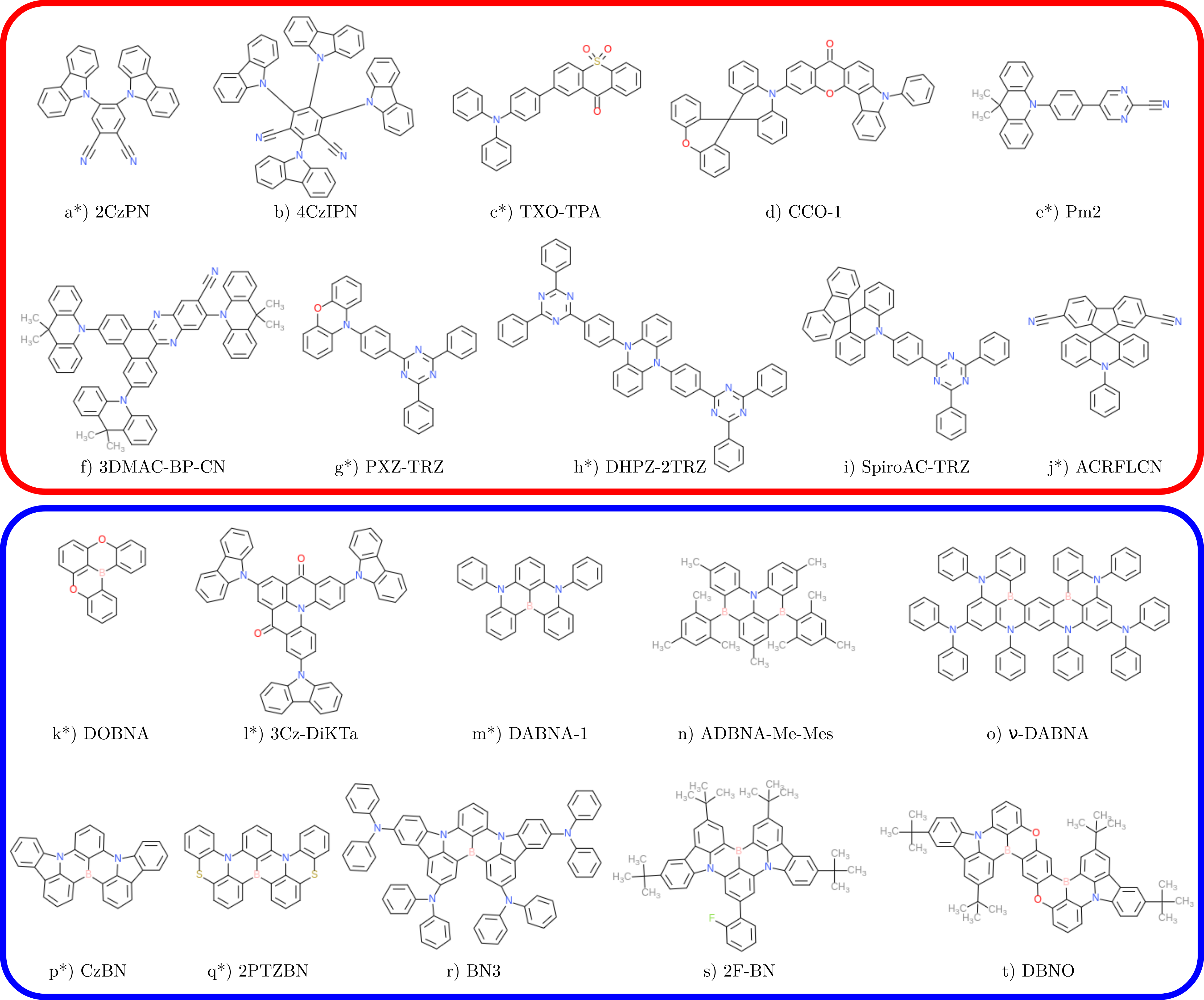}
    \caption{Depiction of the parent molecules, with D-A (MR) molecules in the red (blue) box. The molecules which were used to generate the set of benchmark molecules are labeled with $\ast$. Images were generated using ChemDoodle Web Components~\cite{Burger2015}. \label{fig:parent_molecules}}
\end{figure*}

In order to assess the workflow performance for both D-A and MR molecules, a set of 10 parent TADF emitters was defined for each. These molecules, which are depicted in Fig.~\ref{fig:parent_molecules}, were chosen for diversity of both atomic structure and emission wavelength. The employed SMILES strings and a summary of key properties are given in Tables I and II in the SI. Before running the full workflow, benchmark calculations were performed to calibrate the thresholds for the DFT and TDDFT calculations, as discussed below.

\subsection{Experimental Benchmarks}

\begin{figure*}[tb]
 \centering
 \begin{subfigure}[t]{0.34\textwidth}
    \centering
    \includegraphics[scale=0.54]{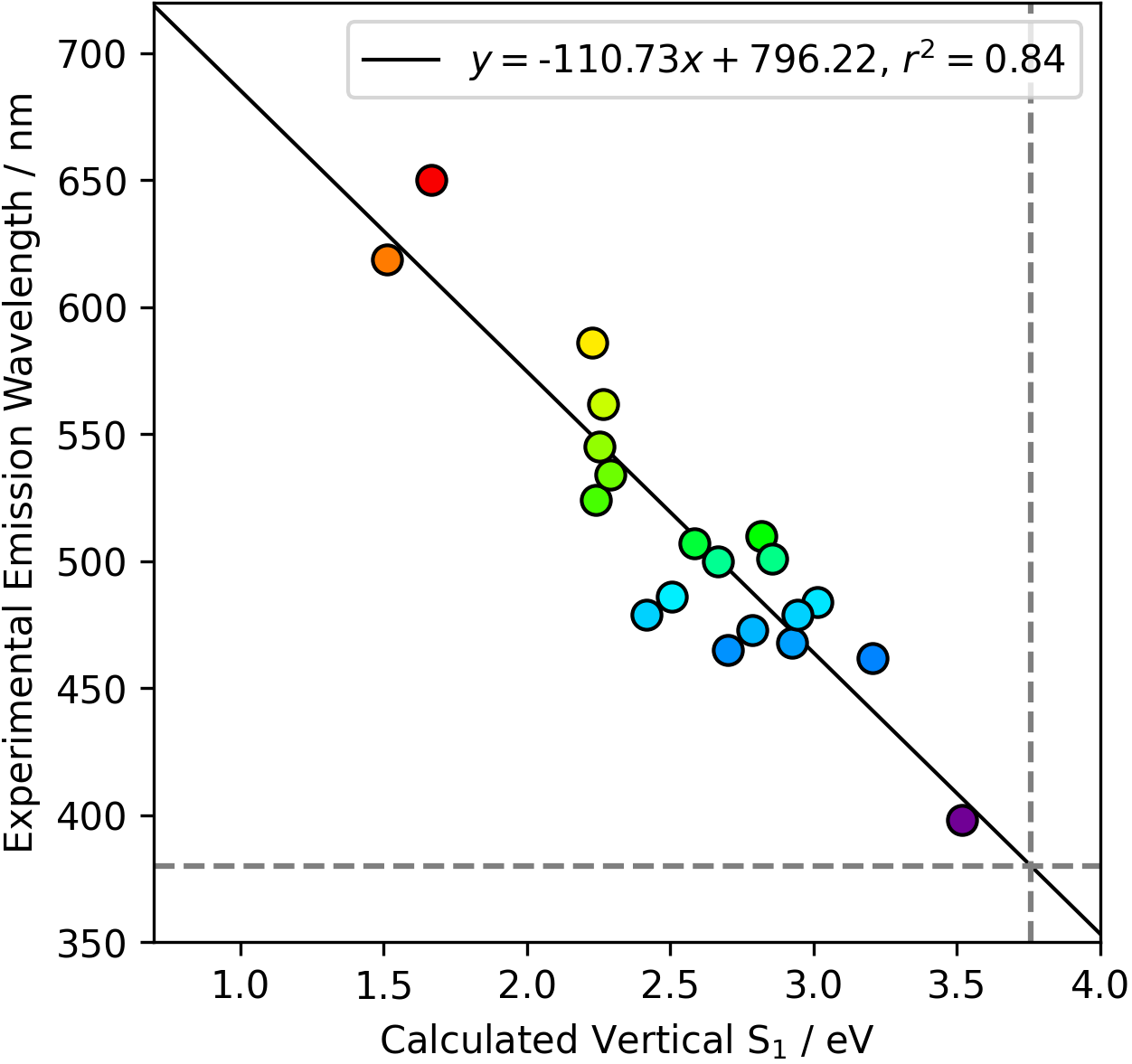}
    \caption{Experimental emission wavelength \emph{vs.}\ calculated S$_1$ energy  for the parent molecules. The dashed lines denote the visible light cutoff and corresponding S$_1$ threshold. }
    \label{fig:parent_s1_benchmark}
\end{subfigure}
\hspace{2pt}
 \begin{subfigure}[t]{0.64\textwidth}
    \centering
    \includegraphics[scale=0.54]{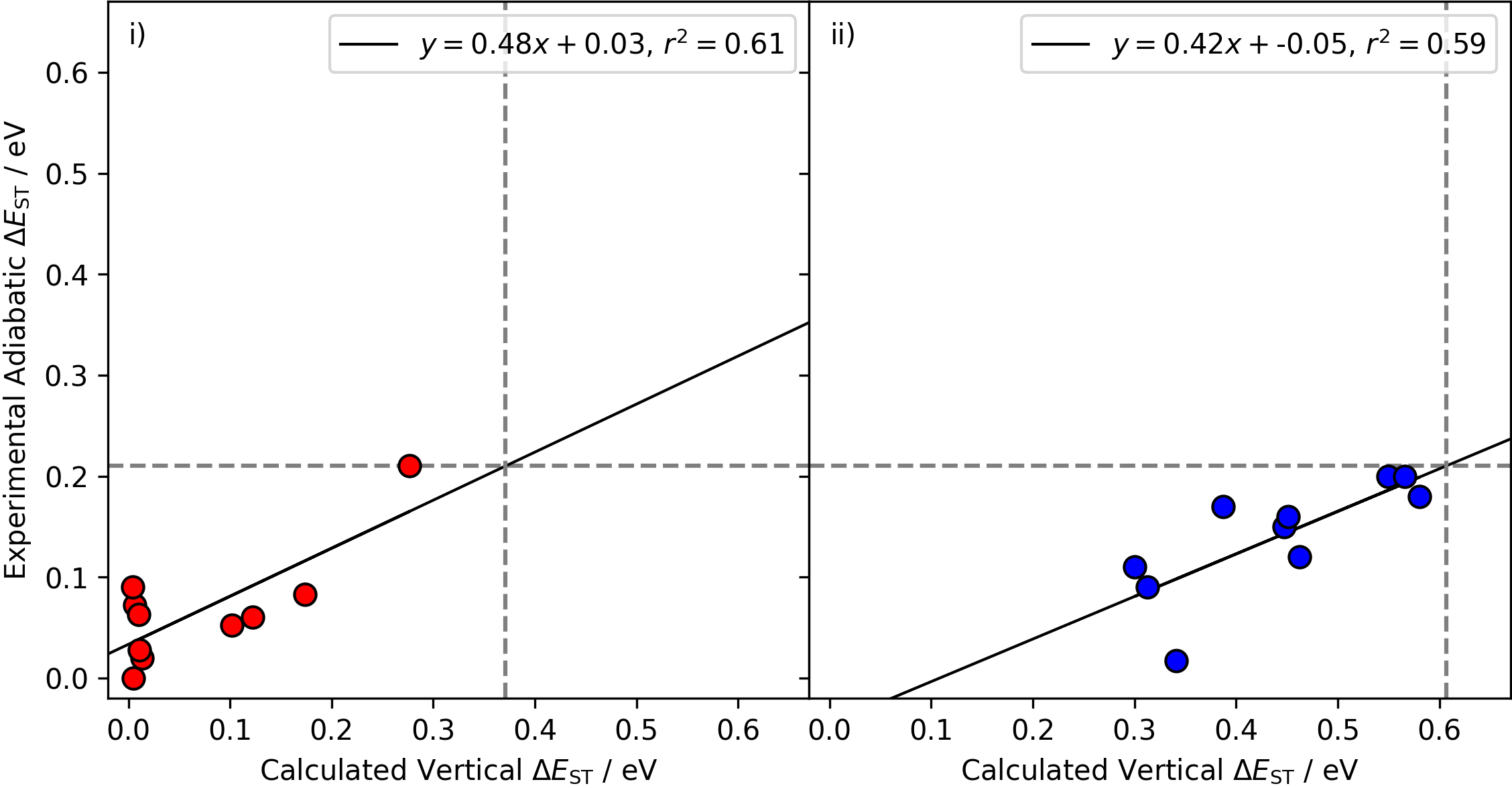}
    \caption{Experimental adiabatic $\Delta E_{\mathrm{ST}}$ energies \emph{vs.}\ calculated vertical $\Delta E_{\mathrm{ST}}$ energies  for i) the D-A parent molecules, and ii) the MR parent molecules. The chosen experimental $\Delta E_{\mathrm{ST}}$ threshold value of 0.21~eV and corresponding calculated threshold value (obtained using the linear fit) are depicted as dashed lines.}
    \label{fig:parent_est_benchmark}
\end{subfigure}
 \begin{subfigure}[t]{0.34\textwidth}
    \centering
    \includegraphics[scale=0.54]{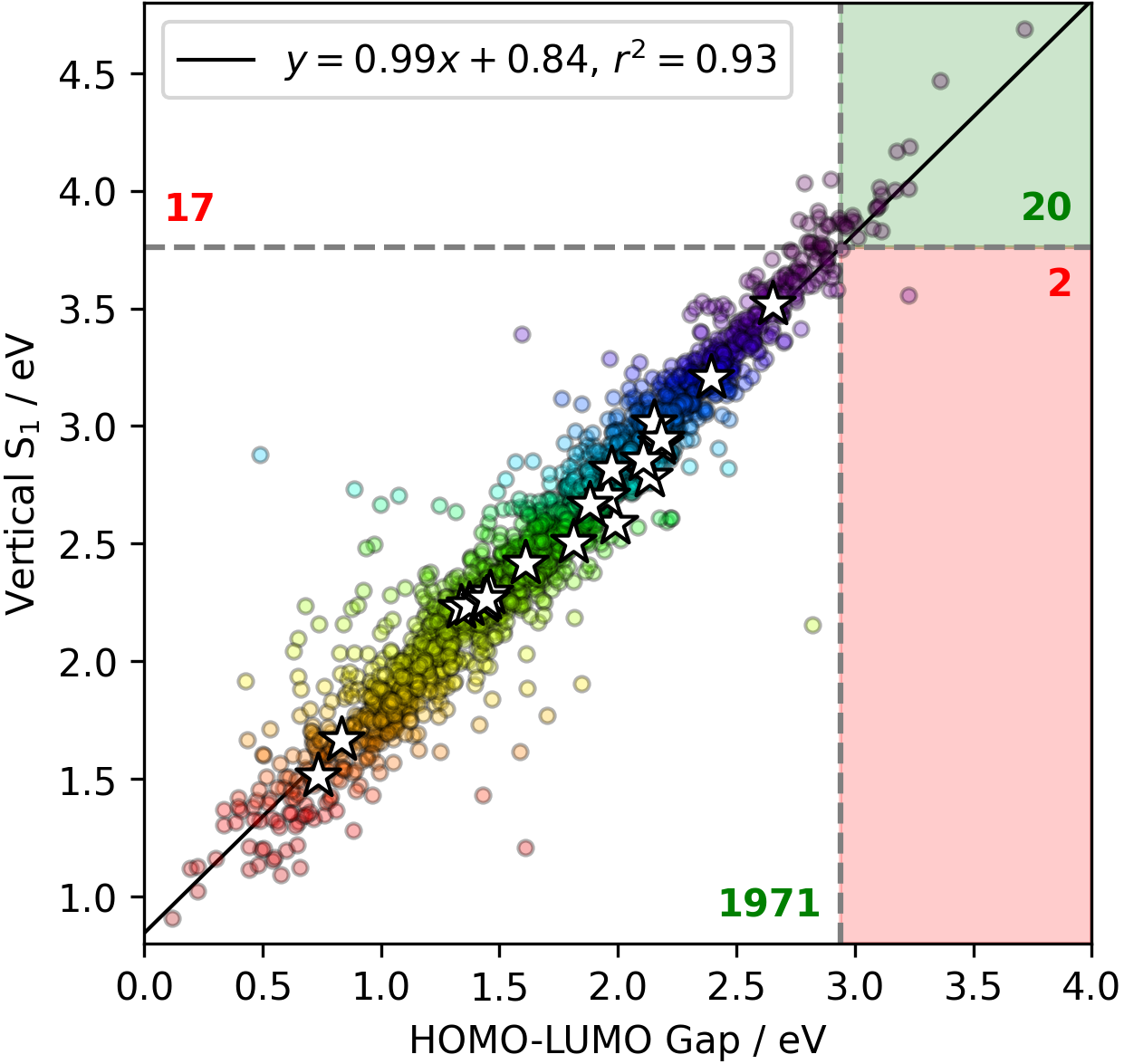}
    \caption{S$_1$ energy \emph{vs.}\ HOMO-LUMO gap for the benchmark molecules. The colour is the predicted emission wavelength based on a), and the dashed lines show the S$_1$ and corresponding HOMO-LUMO gap thresholds.}
    \label{fig:overall_s1_benchmark}
\end{subfigure}
\hspace{2pt}
 \begin{subfigure}[t]{0.64\textwidth}
    \centering
    \includegraphics[scale=0.54]{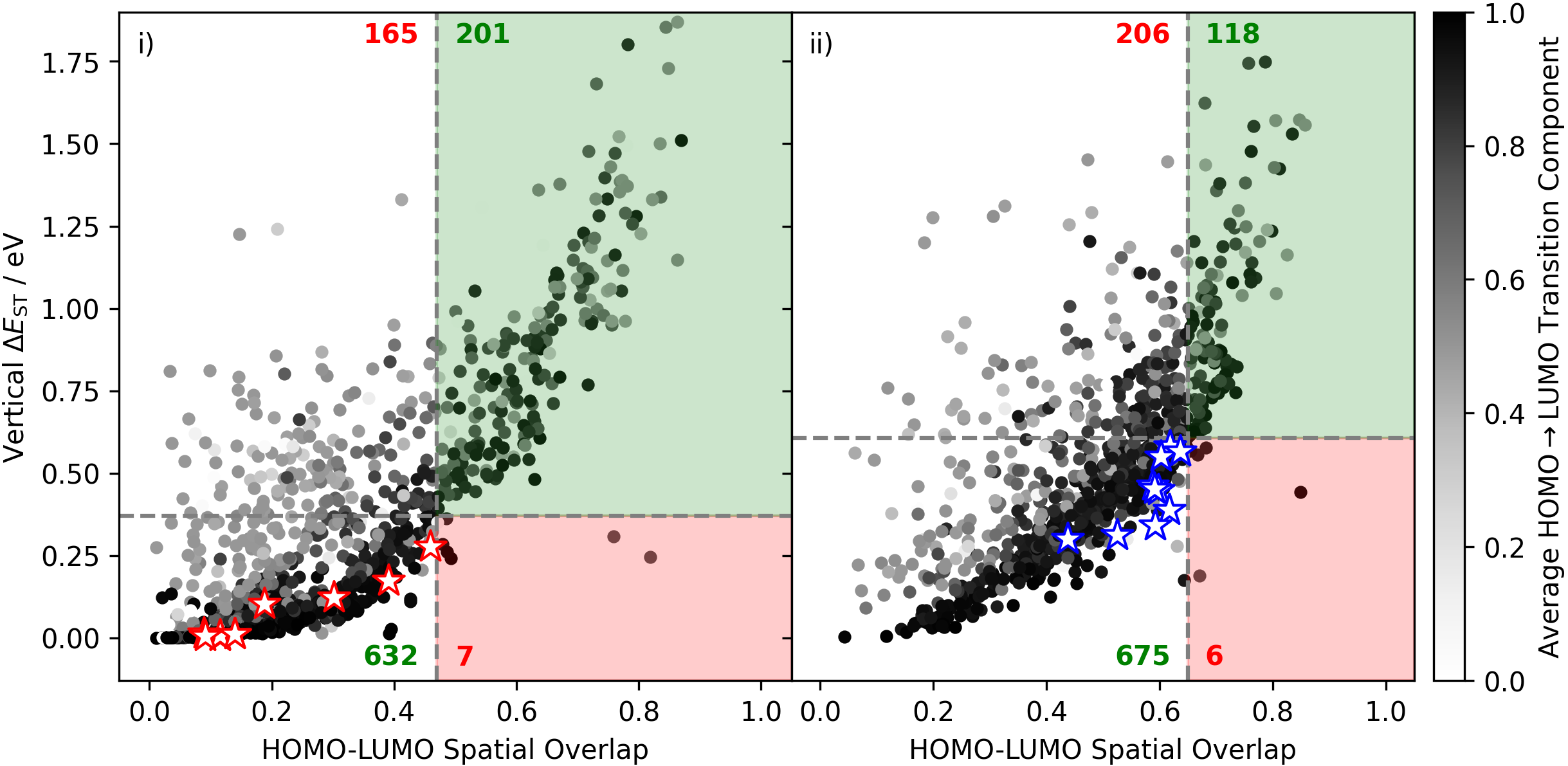}
    \caption{$\Delta E_{\mathrm{ST}}$ \emph{vs.}\ HOMO-LUMO spatial overlap for i) D-A and ii) MR benchmark molecules. The shading represents the average S$_1$/$T_1$ HOMO$\rightarrow$LUMO transition component. The threshold for the HOMO-LUMO spatial overlap was chosen so that no more than 1\% of molecules are incorrectly eliminated based on the threshold values for $\Delta E_{\mathrm{ST}}$ obtained in b), while also ensuring all parent molecules are retained.}
    \label{fig:overall_est_benchmark}
\end{subfigure}
\caption{Benchmarking plots for determining the screening thresholds for HOMO-LUMO gaps and overlaps. 
The green (red) shaded regions in c) and d) show molecules which are correctly (wrongly) eliminated following the HOMO-LUMO calculation. The corresponding numbers of molecules are written in green (red), as are those which are correctly (needlessly) retained in the unshaded areas.
For c) and d) parent molecules are highlighted using the $\bigstar$ symbol. S$_1$, $\Delta E_{\mathrm{ST}}$ and HOMO$\rightarrow$ LUMO transition components were calculated using TDDFT at the B3LYP/6-31G* level on the DFT-optimised geometries, while HOMO-LUMO gaps and overlaps were calculated using PBE and a wavelet basis on the xTB-optimised geometries. Experimental values are from Refs~\cite{Kim2017,Uoyama2012,Wang2014,Fu2023,Pan2016,Huang2021,Tanaka2012,Huang2013,Lee2015,Lin2016,Mehes2012,Ikeda2020,Wu2022,Hatakeyama2016,Oda2019,Kondo2019,Xu2021,Hua2021,Zhang2019,Cai2022}; the $\Delta E_{\mathrm{ST}}$ value for DHPZ-2TRZ is recorded as ``nearly zero'', and so was approximated to zero.}
\label{fig:benchmarks}
\end{figure*}

Fig.~\ref{fig:parent_s1_benchmark} shows a comparison between TDDFT vertical S$_1$ energies and experimental emission wavelengths for the parent molecules. Despite the lack of excited state geometry optimisation and differences with respect to experimental conditions (e.g.\ measurements were performed in various solvents or thin films \emph{vs.}\ gas phase calculations), the correlation between theory and experiment is very good ($r^2=0.84$), so that vertical S$_1$ energies provide a reasonable estimate of emission wavelengths for an HTVS approach. The linear fit shown in Fig.~\ref{fig:parent_s1_benchmark} was therefore used both to predict the emission colour for all subsequent calculations in this work, and for defining a cut-off threshold such that molecules with calculated S$_1$ values giving a predicted emission wavelength below 380~nm can be neglected due to expected emission outside the visible range. No cut-off was defined at longer wavelengths since none of the child molecules in subsequent benchmarks were predicted to exhibit non-visible emission at this end of the spectrum. 

Fig.~\ref{fig:parent_est_benchmark} shows a comparison between calculated and experimental $\Delta E_{\mathrm{ST}}$ values, separated by emitter type. As expected, absolute $\Delta E_{\mathrm{ST}}$ values are not well captured, particularly for MR emitters, however the correlation is similar for both types of emitters ($r^2=0.61$ ($0.59$) for D-A (MR)). While not as accurate as the emission wavelength predictions, the trends in $\Delta E_{\mathrm{ST}}$ are sufficiently well approximated to define a threshold $\Delta E_{\mathrm{ST}}$ value above which molecules can be discarded. Separate D-A and MR thresholds were found using the linear fits and the maximum parent molecule experimental $\Delta E_{\mathrm{ST}}$ value.
While it is possible that some promising molecules will be erroneously discarded using this approach, these are likely to have $\Delta E_{\mathrm{ST}}$ on the higher end of the targeted range, while the cost savings with respect to a more expensive method make this a reasonable compromise within the context of an HTVS approach. 

\subsection{HOMO-LUMO Thresholds}

Having determined TDDFT thresholds, further benchmarks were performed to determine thresholds for the HOMO-LUMO gap and spatial overlap. These were done for the parent molecules, and for child molecules generated from a subset of the parent molecules (see Fig.~\ref{fig:parent_molecules}). For each parent molecule in this subset, single point DFT calculations, ground state DFT geometry optimisations and single point TDDFT calculations were performed for 200 molecules which passed the initial filters.

The correlation between HOMO-LUMO gaps and calculated vertical S$_1$ energies for the benchmark molecules is shown in Fig.~\ref{fig:overall_s1_benchmark}. Semi-local functionals such as PBE are not expected to accurately predict band gaps, however the band gaps themselves are not directly relevant, except as a means for predicting the emission colour. Indeed, as shown in Fig.~\ref{fig:overall_s1_benchmark}, the correlation between the PBE-calculated HOMO-LUMO gap and the TDDFT/B3LYP S$_1$ energy is excellent ($r^2=0.93$).  The linear fit depicted in Fig.~\ref{fig:overall_s1_benchmark} was therefore used to determine a threshold for the HOMO-LUMO gap, using the threshold for calculated S$_1$ values discussed above (see Table III in the SI for values). Also shown in Fig.~\ref{fig:overall_s1_benchmark} is the number of molecules which are correctly (wrongly) retained (discarded) using the HOMO-LUMO gap as a predictor. Only 2 out of 2010 molecules are erroneously discarded, with 17 being needlessly calculated. A corresponding breakdown divided by parent molecule is given in Table IV in the SI. 

Fig.~\ref{fig:overall_est_benchmark} shows a comparison between the HOMO-LUMO spatial overlap and calculated vertical $\Delta E_{\mathrm{ST}}$, with D-A and MR molecules plotted separately. The correlation is much weaker than between the HOMO-LUMO gap and S$_1$ energy, in large part due to the fact that not all S$_1$ and T$_1$ states are dominated by a HOMO$\rightarrow$LUMO transition. Indeed, the majority of the outliers are molecules which have S$_1$ and/or T$_1$ transitions which are not HOMO$\rightarrow$LUMO dominated. Despite the weaker correlation, there are only a handful of molecules which have a relatively high HOMO-LUMO spatial overlap and small $\Delta E_{\mathrm{ST}}$. Therefore, threshold values for the HOMO-LUMO spatial overlap were chosen such that no more than 1\% of molecules with $\Delta E_{\mathrm{ST}}$ below the threshold would be erroneously discarded, while also ensuring all parent molecules would be retained. As shown in Fig.~\ref{fig:overall_est_benchmark}, this leads to only a handful of molecules being incorrectly lost. Although a relatively large number of molecules with high $\Delta E_{\mathrm{ST}}$ are retained, 10\% (6\%) of D-A (MR) molecules are nonetheless correctly discarded, resulting in significant computational savings due to skipping subsequent workflow steps.  Final threshold values and the breakdown of molecules lost/retained divided by parent molecule are given in Tables III and IV in the SI respectively. The use of the HOMO-LUMO gap for predicting $\Delta E_{\mathrm{ST}}$ was also explored, but this showed no correlation, while the use of the HOMO-LUMO overlap $\times$ the HOMO-LUMO gap led to a higher number of molecules with high $\Delta E_{\mathrm{ST}}$ being needlessly retained.

\section{Workflow Results}

Having determined the threshold values, all stages of the workflow were run for each of the 20 parent molecules. In the following the overall performance of the workflow is first discussed, followed by the trends in properties of the potential candidate child molecules, i.e.\ those which passed all filters. The section finishes with a discussion highlighting some of the most interesting candidate child molecules.

\subsection{Workflow Performance}

The relative importance of different filter stages can be assessed \emph{via} funnel plots showing the number of child molecules reaching each stage, an example of which is given for TXO-TPA in Fig.~\ref{fig:funnel} (see Figures 1-4 in the SI for the other parent molecules). First, the initial filters are responsible for eliminating a large number of child molecules: on average roughly a quarter are eliminated due to being ionic, while nearly three quarters have been eliminated by the time synthesisability has been taken into account. This suggests that the random nature of STONED is not very efficient at generating relevant molecules, indeed in the worst cases the generated molecules are small fragments or unrealistic structures. Nonetheless, the initial filters are very cheap, and are able to filter out the majority of the obviously unsuitable molecules. 

The most noticeable variations between parent molecules are due to the SAscore filter. This is partly due to the fact that the larger parent molecules tend to have higher SAscores, e.g.\ DBNO has a high SAscore of 4.26, close to the threshold of 4.5, so that only 86 molecules passed this stage. At the other extreme, 2CzPN, being much smaller and having an SAscore of 2.29, had 653 molecules pass this stage. In future, it might therefore be advisable to vary the threshold depending on the parent molecule, or investigate other methods for predicting synthesisability. In general, MR molecules have fewer child molecules passing all filters, influenced by the larger average molecule size and correspondingly higher SAscores. By the end of the workflow the gap between the number of remaining molecules narrows, with e.g.\ 77, 180 and 358 molecules passing the final filters for DBNO, 2CzPN and TXO-TPA respectively, where the latter is the parent molecule for which the most child molecules passed the final filters. 

The workflow was run independently for each parent molecule. Despite the relatively high similarity of some parent molecules, e.g.\ DBNO and 2F-BN have a Tanimoto similarity of 0.70, there were only six pairs of duplicate child molecules making it through to the first DFT calculation stage; only two of these duplicate pairs made it through the final filters. These all came from D-A parent molecules, with any duplicates generated between MR parent molecules or between D-A and MR parent molecules (primarily small fragments) eliminated at early stages. In future a database approach could be used to eliminate molecules which have already been generated. Another minor limitation of the current implementation is that STONED occasionally generates molecules which are identical to the parent molecule. When such molecules had identical SMILES strings they were filtered out, but there were a few instances of molecules with different SMILES strings which nonetheless had a Tanimoto similarity of 1. In some cases the calculated properties were slightly different from those of the parent molecule due to small differences in the geometry, highlighting the dependency on the initial SMILES string.

One test for evaluating the effectiveness of an HTVS workflow is if existing emitters are rediscovered. To this end, the final child molecules were compared to a set of around 40 known TADF emitters, chosen due to structural similarities with the parent molecules. Only one such molecule was rediscovered: Py2 (Pm2 child molecule \textbf{114}), which is somewhat surprising, particularly in cases where these known emitters have high similarity with respect to the related parent molecule, e.g.\ 4CzIPN has a Tanimoto similarity of 0.98 with respect to 4CzTPN, but was nonetheless not rediscovered. 
One possible explanation is that a single mutation is insufficient to explore a diverse enough chemical space, however repeating the library generation step for increasing numbers of mutations, with 10,000 molecules generated each for 1 through to 10 mutations (i.e.\ generating 100,000 unique SMILES per parent molecule) did not result in any further rediscoveries. Furthermore, the average Tanimoto similarity with respect to the parent molecule drops off rapidly with the number of mutations, so that in most cases by 4-5 mutations the average Tanimoto similarity with respect to the parent molecule is less than 0.25. Thus while increasing the number of mutations increases diversity, this rapidly tends towards a region of chemical space which is far from the parent molecule. At the same time, the average size of the child molecules tends to decrease with increasing number of mutations, while some of the existing emitters are much larger than the most similar parent molecule, and thus seemingly out of reach of the STONED algorithm. Similarly, many of the existing emitters have high symmetry, which a random approach tends to break. It might therefore be interesting in future to combine a random approach with e.g.\ a genetic algorithm or ML-based approach to target the mutations more effectively. However, as shown below, the workflow resulted in a number of child molecules which have promising properties, thus it was nonetheless successful in exploring a relevant and useful region of chemical space around the parent molecules.

\subsection{Trends in Candidate Child Molecule Properties}\label{sec:child_trends}

\begin{figure}[ht]
    \centering
    \includegraphics[scale=0.5]{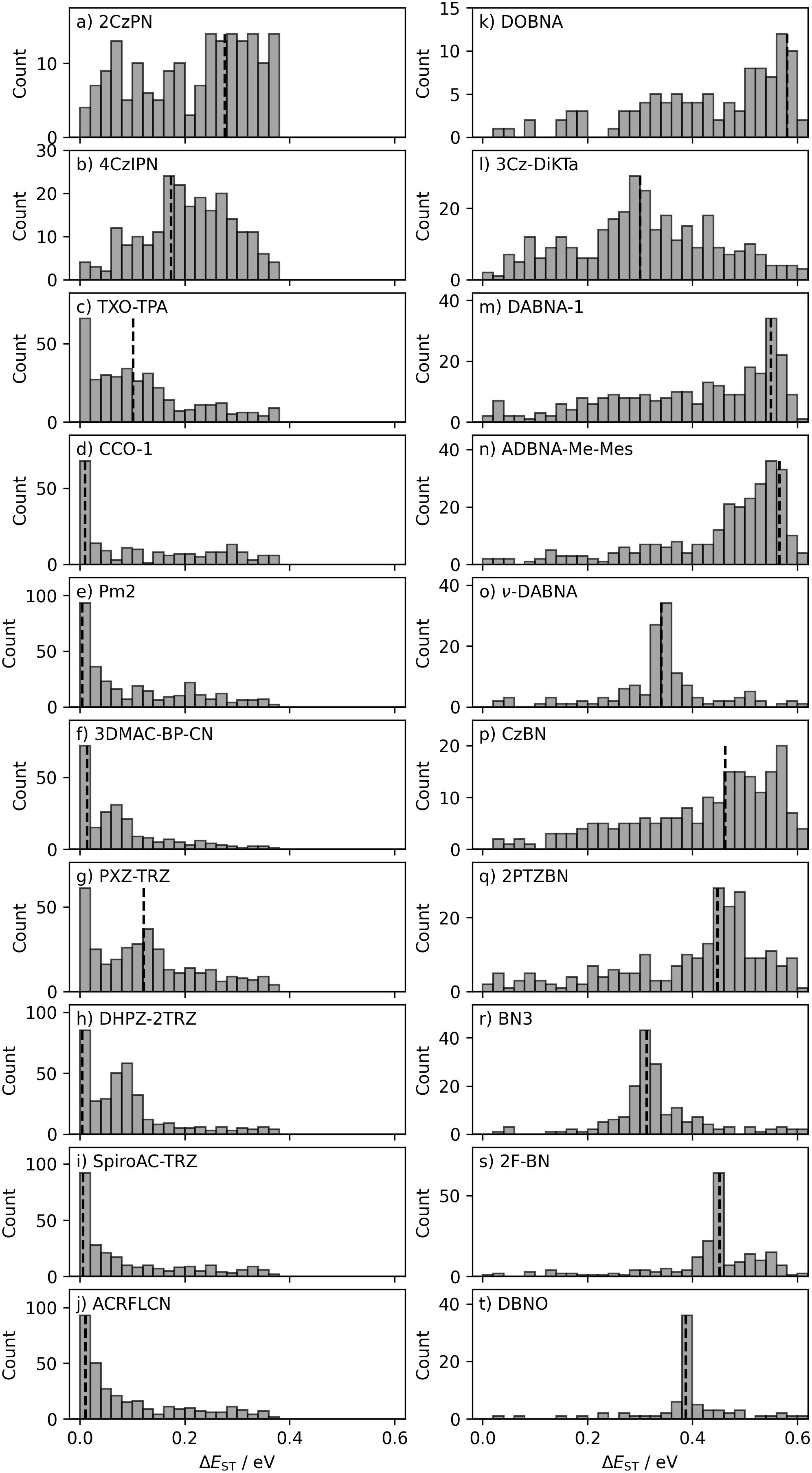}
    \caption{$\Delta E_{\mathrm{ST}}$ values for molecules which passed all workflow filters. Values for parent molecules are depicted with a dashed vertical line.
    \label{fig:est_final}}
\end{figure}

The distribution of $\Delta E_{\mathrm{ST}}$ values for child molecules which passed all filters is shown in Fig.~\ref{fig:est_final}. Most of the parent molecules have peaks around the $\Delta E_{\mathrm{ST}}$ of the parent molecule, with e.g.\ 2F-BN and DBNO having very sharp peaks, while other molecules like 2CzPN have a wider spread of values. The majority of D-A parent molecules have a peak in $\Delta E_{\mathrm{ST}}$ values close to 0~eV. There are two exceptions: 2CzPN and 4CzIPN have notably fewer molecules with very small $\Delta E_{\mathrm{ST}}$. This in line with the higher parent $\Delta E_{\mathrm{ST}}$ values, where 2CzPN also has a high experimental $\Delta E_{\mathrm{ST}}$ value, while that of 4CzIPN is on the higher end and overestimated with respect to experiment. Nonetheless both molecules still have a number of molecules with small $\Delta E_{\mathrm{ST}}$. Unsurprisingly the MR child molecules typically have larger calculated $\Delta E_{\mathrm{ST}}$ values, due to both higher experimental values and TDDFT overestimating. Nonetheless, for each MR parent molecule the child molecules show a range of $\Delta E_{\mathrm{ST}}$ values, including very small values. There is also a spread in HOMO-LUMO spatial overlap values (Figure 5 in the SI). Although only applicable for HOMO-LUMO dominated excitations, it is interesting to note that there are MR child molecules with very small overlap, indicating possible D-A character. Thus while the majority of child molecules are of the same character as their parent, the random nature of STONED allows for the possibility of generating D-A molecules from MR parents and \emph{vice versa}, as well as molecules which may have aspects of both, which could be explored further in future work.
                                                                        
There were also a wide range of S$_1$ oscillator strengths across the final child molecules (see Figures 6-9 in the SI). Slightly different trends were seen when considering $\Delta E_{\mathrm{ST}}$ \emph{vs.}\ $f_{S_1}$ for the D-A molecules compared to the MR molecules, with a greater number of D-A child molecules having very small $f_{S_1}$, in line with a number of D-A parent molecules also having very small $f_{S_1}$ (Figure 10                                                                                                                                                                                                                                                                                                                                                                                                                                                                                                                                                                                                                                                                                                                                                                                                                                                                                                                                                                                                                                                                                                                                                                                                                                                                                                                                                                                                                                                                                                                                                                                                                                                                                                                                                                                                                                                                                                                                                                                                                                                                                                                                                                                                                                                                                                                                                                                                                                                                                                                                                                                                                                                                                                                                                                                                                                                                                                                                                                                                                                                                                                                                                                                                                                                                                                                                                                                                                                                                                                                                                                                                                                                                                                                                                                                                                                                                                                                                                                                                                 in the SI). However, in both cases, as is to be expected $\Delta E_{\mathrm{ST}}$ and $f_{S_1}$ are in competition, with the molecules having the smallest $\Delta E_{\mathrm{ST}}$ also having smaller $f_{S_1}$ values. For this reason, and because of the very small $f_{S_1}$ values of some parent molecules, it was decided to not filter molecules based on a particular $f_{S_1}$ threshold, and instead balance both factors when considering candidate child molecules. Given the differences in behaviour for different parent molecules, this was easier than e.g.\ using a combined function of merit, as has been done in previous work~\cite{Shu2015}.

\begin{figure}[ht]
    \centering
    \includegraphics[scale=0.5]{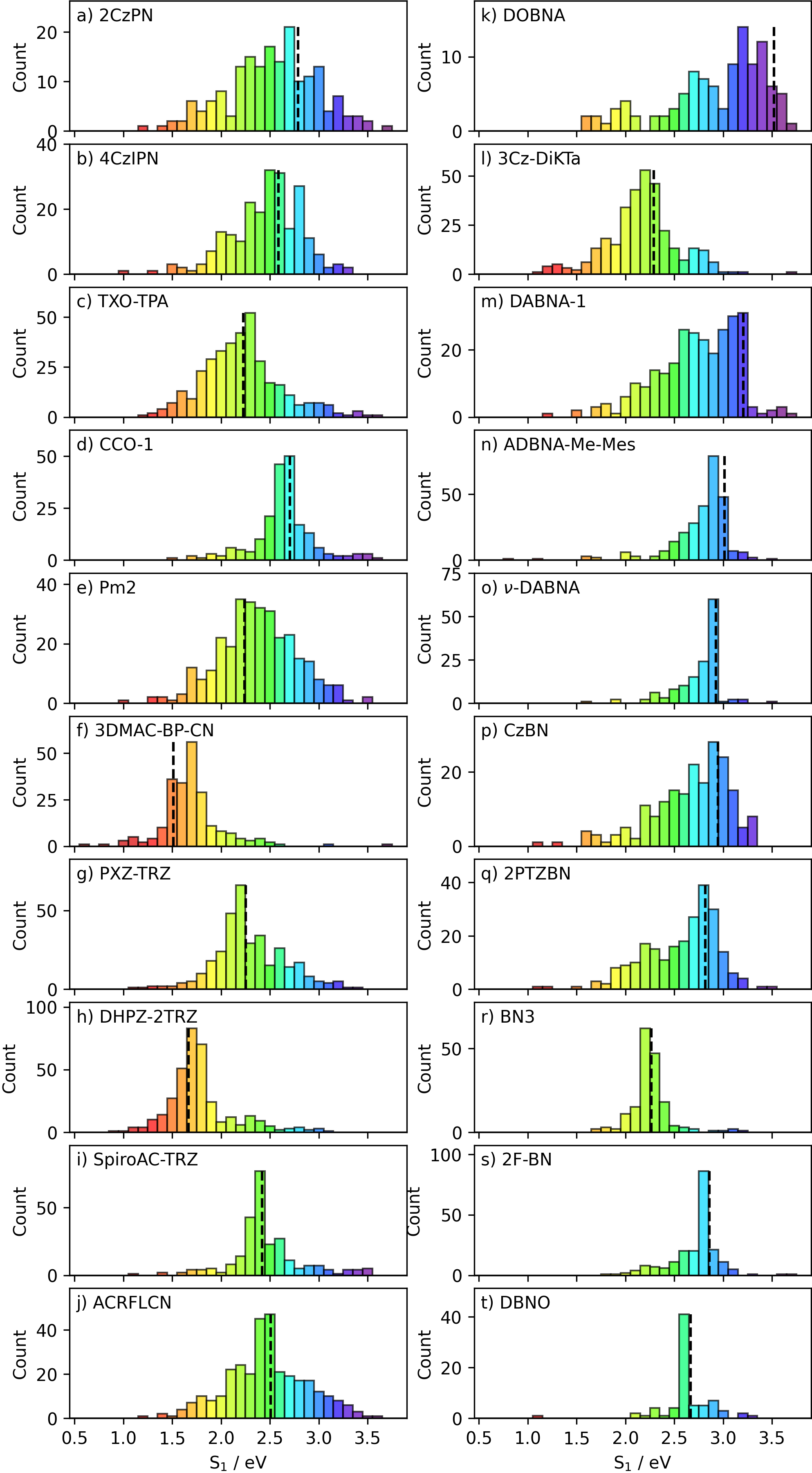}
    \caption{Vertical S$_1$ values for molecules which passed all workflow filters, with corresponding predicted emission colour. Values for parent molecules are depicted with a dashed vertical line.
    \label{fig:s1_final}}
\end{figure}

The final distribution of (vertical) S$_1$ energies is shown in Fig.~\ref{fig:s1_final}, including the predicted emission colours. In most cases there is a peak around the parent molecule value, although some have very sharp peaks, e.g.\ BN3 and DBNO, while others like 2CzPN have more even distributions of S$_1$ energies. Although there are limitations to this approach to predicting emission colour, it is nonetheless clear that some parent molecules more easily lend themselves to colour tuning across the full spectrum. Indeed, most of the MR parent molecules tend to have a narrower spread of values, with relatively few red emitters, although this is also highly influenced by the distribution of emission colours of the MR parent molecules themselves, as well as a number of the predicted emission colours for the MR parent molecules being blue-shifted compared to experiment.

\subsection{Candidate Child Molecules}

We conclude the results by discussing interesting candidate child molecules for each parent. This includes extremes of emission colour, small $\Delta E_{\mathrm{ST}}$, and balancing small $\Delta E_{\mathrm{ST}}$ with reasonable $f_{\mathrm{S}_1}$. Where applicable, we also discuss any notable structural features. Child molecules are indexed separately for each parent molecule.

\subsubsection{D-A Molecules.}

\paragraph{2CzPN.}

There is a broad spread of predicted emission colours for the candidate child molecules, although most are blue or green, in line with 2CzPN's blue emission. At the extremes, there are four red/orange-red candidates; of these only \textbf{1392} has a small $\Delta E_{\mathrm{ST}}$ (0.09~eV), below that of 2CzPN. There are several violet emitters, with three having $\Delta E_{\mathrm{ST}}<0.2$~eV (\textbf{1583}, \textbf{1657} and \textbf{1723}), although all have $f_{\mathrm{S}_1}\leq 0.001$. A total of 38 molecules have $\Delta E_{\mathrm{ST}}<0.1$~eV, of which 14 are below 0.05~eV, covering a broad range of colours. Balancing $\Delta E_{\mathrm{ST}}$ and $f_{\mathrm{S}_1}$, there are no molecules with $\Delta E_{\mathrm{ST}}<0.2$~eV and $f_{\mathrm{S}_1}>0.1$, although 2CzPN itself has $f_{\mathrm{S}_1}=0.05$. There are however six candidates with $\Delta E_{\mathrm{ST}}<0.2$~eV and $f_{\mathrm{S}_1}>0.05$, which are all blue or green emitters. Twelve molecules have smaller $\Delta E_{\mathrm{ST}}$ and larger $f_{\mathrm{S}_1}$ than 2CzPN, with a wide range of Tanimoto similarities with respect to the parent molecule and blue through to yellow emission. Both S and I heteroatoms feature in these molecules.

\paragraph{4CzIPN.}

Many child molecules are green emitters, like 4CzIPN itself. There are two red emitters (\textbf{1277} and \textbf{1690}), where \textbf{1277} has $\Delta E_{\mathrm{ST}}<0.1$~eV but also small $f_{\mathrm{S}_1}$ (0.003), as well as a few orange emitters. There are various violet and deep blue emitters, but many of these have larger $\Delta E_{\mathrm{ST}}$ than 4CzIPN, with \textbf{772} being the shortest wavelength emitter (459~nm) that has smaller $\Delta E_{\mathrm{ST}}$; it also has $f_{\mathrm{S}_1}=0.06$, just below the parent molecule. There are 29 molecules with $\Delta E_{\mathrm{ST}}<0.1$~eV, of which seven have $\Delta E_{\mathrm{ST}}<0.05$~eV, spanning blue to red emission. Two molecules have $\Delta E_{\mathrm{ST}}<0.2$~eV and $f_{\mathrm{S}_1}>0.1$ (\textbf{193} and \textbf{1574}), which are both blue. Seven molecules have smaller $\Delta E_{\mathrm{ST}}$ and larger $f_{\mathrm{S}_1}$ than the parent molecule, with blue to green emission, and Tanimoto similarities with respect to 4CzIPN ranging from 0.63 to 0.91. Some of these have lost functional groups compared to 4CzIPN. P and Cl heteroatoms feature, with S, Br and O heteroatoms also appearing among molecules with small $\Delta E_{\mathrm{ST}}$.

\paragraph{TXO-TPA.}

There is a broad peak of predicted emission colours around green, in line with the predicted emission colour of TXO-TPA, whose emission is predicted to be green instead of yellow. There are various red and dark orange emitters, including three with predicted wavelengths above 640~nm (\textbf{438}, \textbf{569} and \textbf{974}), all with small $\Delta E_{\mathrm{ST}}$ but also very small $f_{\mathrm{S}_1}$ ($<0.001$). There are also various violet and deep blue emitters, including five with predicted wavelengths under 420~nm. Of these, both \textbf{248} and \textbf{529} have $\Delta E_{\mathrm{ST}}$ values smaller than TXO-TPA, but they also have smaller $f_{\mathrm{S}_1}$ than the parent. There are 186 molecules with $\Delta E_{\mathrm{ST}}<0.1~$eV, of which 109 have $\Delta E_{\mathrm{ST}}$<0.05~eV, ranging from blue to red,. Three molecules have $\Delta E_{\mathrm{ST}}<0.2$~eV and $f_{\mathrm{S}_1}>0.1$ (\textbf{748}, \textbf{1423} and \textbf{1849}), which are various shades of green. A number of heteroatoms feature, in particular P, which is also present in two of the six molecules which have both $\Delta E_{\mathrm{ST}}$ and larger $f_{\mathrm{S}_1}$ than TXO-TPA. These molecules are also either green or yellow, and have a range of Tanimoto similarities with respect to TXO-TPA.

\paragraph{CCO-1.}

There is a sharp peak of emitters with green or blue-green emission, in line with the emission colour of CCO-1, which is predicted to be blue-green, somewhat redshifted compared to the blue experimental emission. There are no predicted red emitters and few orange, including two with emission around 600~nm (\textbf{1040} and \textbf{1054}), and one at 619~nm (\textbf{193}). The latter has small $\Delta E_{\mathrm{ST}}$ (0.001~eV) but also small $f_{\mathrm{S}_1}$ ($4\times 10^{-4}$), although this is larger than the very small $f_{\mathrm{S}_1}$ of the parent molecule ($10^{-5}$). There are various violet and deep blue emitters, including seven with emission wavelengths below 420~nm, but none have $\Delta E_{\mathrm{ST}}<0.1$~eV, and those below 0.2~eV have small $f_{\mathrm{S}_1}$. There are 105 molecules with $\Delta E_{\mathrm{ST}}<0.1$~eV, of which 85 have $\Delta E_{\mathrm{ST}}<0.05$~eV, spanning blue to orange emission. One molecule has $\Delta E_{\mathrm{ST}}<0.2$~eV and $f_{\mathrm{S}_1}>0.1$ (\textbf{108}), with predicted green emission. Seventeen molecules have both smaller $\Delta E_{\mathrm{ST}}$ and larger $f_{\mathrm{S}_1}$ than the parent molecule, spanning blue to orange. These molecules mostly have Tanimoto similarities above 0.7 compared to CCO-1, while P, Cl and S heteroatoms all feature.

\paragraph{Pm2.}

There is a broad peak in emission energies around green, in line with Pm2's predicted green emission, which is red-shifted compared to experiment. At the red end of the spectrum, five molecules have predicted wavelengths above 630~nm, all with $\Delta E_{\mathrm{ST}}<0.1$~eV, with \textbf{1746} having the highest $f_{\mathrm{S}_1}$ (0.04), although all are higher than Pm2's very small $f_{\mathrm{S}_1}$ ($2\times 10^{-5}$). There are two violet outliers (\textbf{266} and \textbf{844}), where \textbf{844} has very small $\Delta E_{\mathrm{ST}}$ ($<0.01$~eV), but also small $f_{\mathrm{S}_1}$ ($2\times 10^{-4}$). There are 141 molecules with $\Delta E_{\mathrm{ST}}<0.05$~eV, spanning violet to red, and 34 with $0.05\leq\Delta E_{\mathrm{ST}}<0.1$~eV. There are 15 molecules with $\Delta E_{\mathrm{ST}}<0.2$~eV and $f_{\mathrm{S}_1}>0.1$, and 15 which have both smaller $\Delta E_{\mathrm{ST}}$ and larger $f_{\mathrm{S}_1}$ than Pm2 again spanning a range of emission colours, and with a range of Tanimoto similarities with respect to Pm2. P heteroatoms are present in five of the fifteen, while a range of other heteroatoms feature among the molecules with small $\Delta E_{\mathrm{ST}}$.

\paragraph{3DMAC-BP-CN.}

Most child molecules are red or yellow, in line with the orange-red emission of the parent. This includes two molecules with predicted emission wavelengths above 700~nm (\textbf{36} and \textbf{382}), where \textbf{36} has the smaller $\Delta E_{\mathrm{ST}}$ of the two but also smaller $f_{\mathrm{S}_1}$. There is a single blue (\textbf{1966}) and a single violet molecule (\textbf{268}), both of which are much smaller molecules than 3DMAC-BP-CN, while also having larger $\Delta E_{\mathrm{ST}}$ and small $f_{\mathrm{S}_1}$. There are 165 molecules with $\Delta E_{\mathrm{ST}}<0.1$~eV, of which 98 have $\Delta E_{\mathrm{ST}}<0.05$~eV, spanning green to red emission. There are 11 molecules with $\Delta E_{\mathrm{ST}}<0.2$~eV and $f_{\mathrm{S}_1}>0.1$, and 20 which have both smaller $\Delta E_{\mathrm{ST}}$ and larger $f_{\mathrm{S}_1}$ than 3DMAC-BP-CN, where most of the latter have Tanimoto similarities above 0.7 with respect to the parent molecule, while a range of heteroatoms feature.

\paragraph{PXZ-TRZ.}

In line with PXZ-TRZ's predicted green emission, which is in excellent agreement with experiment, there is a spread of emission colours around green. There are various red and orange emitters, where seven have predicted emission around 630~nm or above. Three of these have smaller $\Delta E_{\mathrm{ST}}$ than PXZ-TRZ, but with $f_{\mathrm{S}_1}<10^{-4}$. There are various blues and violets, including six with predicted emission below 440~nm, of which only \textbf{1617} has $\Delta E_{\mathrm{ST}}$ smaller than PXZ-TRZ, with $f_{\mathrm{S}_1}=0.02$. There are 93 molecules with $\Delta E_{\mathrm{ST}}<0.05$~eV, with blue through to red emission, and 54 with $0.05\leq\Delta E_{\mathrm{ST}}<0.1$~eV. There are 73 molecules with $\Delta E_{\mathrm{ST}}<0.2$~eV and $f_{\mathrm{S}_1}>0.1$, and two with both smaller $\Delta E_{\mathrm{ST}}$ and larger $f_{\mathrm{S}_1}$ than PXZ-TRZ (\textbf{664} and \textbf{1831}), which are green and yellow, and have high similarities with respect to the parent molecule (0.7-0.8). Like many of the emitters with small $\Delta E_{\mathrm{ST}}$, they both contain heteroatoms, in this case B and S.

\paragraph{DHPZ-2TRZ.}

The majority of child molecules are orange or yellow, in line with the predicted orange emission of DHPZ-2TRZ, which is blue-shifted compared to the red of experiment. There are many red emitters, including e.g.\ five with predicted emission wavelengths above 670~nm; of these \textbf{681} has the smallest $\Delta E_{\mathrm{ST}}$ (0.01~eV), but also with very small $f_{\mathrm{S}_1}$ ($<10^{-4}$), which is nonetheless still higher than the very small $f_{\mathrm{S}_1}$ of DHPZ-2TRZ ($10^{-8}$). There are no violet but a few blue emitters, including four at 460~nm or above, although all have $\Delta E_{\mathrm{ST}}\geq0.28$~eV. There are 124 molecules with $\Delta E_{\mathrm{ST}}<0.05$~eV, spanning green to red, and 125 with $0.05\leq\Delta E_{\mathrm{ST}}<0.1$~eV. There are 85 molecules with $\Delta E_{\mathrm{ST}}<0.2$~eV and $f_{\mathrm{S}_1}>0.1$, and 51 which have both smaller $\Delta E_{\mathrm{ST}}$ and larger $f_{\mathrm{S}_1}$ than DHPZ-2TRZ, which are green-yellow to red, and mostly have high Tanimoto similarities to the parent molecule. Some of these feature heteroatoms, though the majority have other structural changes, e.g.\ modified or broken rings.

\paragraph{SpiroAC-TRZ.}

Most child molecules are green, in line with the predicted green emission of the parent molecule, which differs from the blue experimental emission. There is one red outlier (\textbf{284}), which has high $f_{\mathrm{S}_1}$ (0.13), and reasonably small $\Delta E_{\mathrm{ST}}$ (0.16~eV), albeit higher than that of SpiroAC-TRZ. There are also a number of orange emitters e.g.\ two with predicted emission wavelengths of 630~nm or above (\textbf{701} and \textbf{1265}). Of these, \textbf{701} is more promising, having both smaller $\Delta E_{\mathrm{ST}}$ and larger $f_{\mathrm{S}_1}$ than the parent. There are many blue and violet emitters, e.g.\ thirteen with predicted emission wavelengths below 430~nm, which have a range of $\Delta E_{\mathrm{ST}}$ values, but all larger than SpiroAC-TRZ. There are 168 molecules with $\Delta E_{\mathrm{ST}}<0.1$~eV, of which 132 have $\Delta E_{\mathrm{ST}}<0.05$~eV. Four molecules have $\Delta E_{\mathrm{ST}}<0.2$~eV and $f_{\mathrm{S}_1}>0.1$, spanning green to red, and 26 molecules have both smaller $\Delta E_{\mathrm{ST}}$ and larger $f_{\mathrm{S}_1}$ than SpiroAC-TRZ, with a range of Tanimoto similarities with respect to the parent molecule. A range of heteroatoms feature among the more promising child molecules.

\paragraph{ACRFLCN.}

There is a broad spread of predicted emission colours, with green being the most common, in line with the predicted green emission of ACRFLCN, which differs from the blue of experiment. There is one red (\textbf{420}) and three dark orange (\textbf{269}, \textbf{486} and \textbf{1561}) emitters, all with larger $\Delta E_{\mathrm{ST}}$ than the parent. There are many blue and violet, e.g.\ ten with predicted emission wavelengths below 430~nm, two of which have $\Delta E_{\mathrm{ST}}<0.1$~eV (\textbf{990} and \textbf{1884}). There are 156 molecules with $\Delta E_{\mathrm{ST}}<0.05$~eV, spanning a range of colours, and 50 with $0.05\leq\Delta E_{\mathrm{ST}}<0.1$~eV. No molecules have $\Delta E_{\mathrm{ST}}<0.2$~eV and $f_{\mathrm{S}_1}>0.1$, or even $\Delta E_{\mathrm{ST}}<0.2$~eV and $f_{\mathrm{S}_1}>0.05$, with most child molecules having small $f_{\mathrm{S}_1}$, in common with the small $f_{\mathrm{S}_1}$ of ACRFLCN ($4\times 10^{-4}$). There are 15 molecules with smaller $\Delta E_{\mathrm{ST}}$ and larger $f_{\mathrm{S}_1}$ than ACRFLCN, which are various shades of green, and have a range of Tanimoto similarities with respect to ACRFLCN. A range of heteroatoms feature in the molecules with small $\Delta E_{\mathrm{ST}}$.

\subsubsection{MR Molecules.}

\paragraph{DOBNA.}

Most child molecules are towards the violet end of the emission spectrum, like DOBNA, but there is still a spread of predicted emission wavelengths, including four molecules above 600~nm. Of these only \textbf{1152} has $\Delta E_{\mathrm{ST}}<0.1$~eV, it also has reasonable $f_{\mathrm{S}_1}=0.04$, albeit smaller than that of DOBNA. A further three molecules have $\Delta E_{\mathrm{ST}}<0.1$~eV (\textbf{248}, \textbf{835} and \textbf{1303}), while eight more have $0.1\leq\Delta E_{\mathrm{ST}}<0.2$~eV. These have a range of $f_{\mathrm{S}_1}$ values, with \textbf{1152} having the largest, and range from violet to orange. There are no molecules with $\Delta E_{\mathrm{ST}}<0.3$~eV and $f_{\mathrm{S}_1}>0.1$, and only one with $\Delta E_{\mathrm{ST}}<0.3$~eV and $f_{\mathrm{S}_1}>0.05$: \textbf{1374}, which is predicted to be violet. There are five molecules with both smaller $\Delta E_{\mathrm{ST}}$ and larger $f_{\mathrm{S}_1}$ than DOBNA, all with blue/violet emission, and most of which have high similarity to the parent molecule, with e.g.\ the only change being the addition/substitution of a a single heteroatom.

\paragraph{3Cz-DiKTa.}

While there is a peak around the green emission of 3Cz-DiKTa, there is a range of emission colours. There are a number of red and orange emitters, with e.g.\ 13 molecules with predicted emission wavelengths above 630~nm, several of which have small $\Delta E_{\mathrm{ST}}$. There is one violet outlier (\textbf{289}), which is much smaller than the parent molecule, plus one other (\textbf{874}) with emission below 450~nm. Both have larger $\Delta E_{\mathrm{ST}}$ and smaller $f_{\mathrm{S}_1}$ than the parent molecule.  There are 27 molecules with $\Delta E_{\mathrm{ST}}<0.1$~eV, of which \textbf{323} has the largest $f_{\mathrm{S}_1}$ (0.02) and 42 molecules with $0.1\leq\Delta E_{\mathrm{ST}}<0.2$~eV. Three molecules have both $\Delta E_{\mathrm{ST}}<0.3$~eV and $f_{\mathrm{S}_1}>0.1$ (\textbf{371}, \textbf{1904} and \textbf{1974}), which lie in the green to yellow emission range, where \textbf{371} is interesting in that the only difference is the replacement of the N in the DiKTa core with a B atom. There are 13 molecules which have both smaller $\Delta E_{\mathrm{ST}}$ and larger $f_{\mathrm{S}_1}$ than 3Cz-DiKTa, which have green to orange emission, and Tanimoto similarites with respect to the parent molecule of between 0.5 to 0.75.

\paragraph{DABNA-1.}

Most child molecules are blue or green, in line with the deep blue emission of DABNA-1.  There is one red molecule (\textbf{1577}) and a further five with predicted emission above 600~nm, all of which have $\Delta E_{\mathrm{ST}}$ smaller than the parent. Of these, \textbf{312} and \textbf{855} have an S heteroatom bonded to one of the N atoms, and have $f_{\mathrm{S}_1}$ around 0.05. Four molecules have predicted emission below 400~nm, all with smaller $\Delta E_{\mathrm{ST}}$ than DABNA-1, and two of which have $f_{\mathrm{S}_1}>0.05$ (\textbf{1450} and \textbf{1777}). There are 37 molecules with $\Delta E_{\mathrm{ST}}<0.2$~eV of which 14 have $\Delta E_{\mathrm{ST}}<0.1$~eV. A number of these contain either P or B atoms. Five molecules have $\Delta E_{\mathrm{ST}}<0.3$~eV and $f_{\mathrm{S}_1}>0.1$, which are all blue or green. Sixteen molecules have smaller $\Delta E_{\mathrm{ST}}$ and higher $f_{\mathrm{S}_1}$ than DABNA-1, all with predicted blue emission and all but two with a Tanimoto similarity below 0.6 with respect to DABNA-1.

\paragraph{ADBNA-Me-Mes.}

The majority of child molecules of ADBNA-Me-Mes are blue or green, in line with the blue emission of the parent molecule.  There are two predicted red emitters (\textbf{845} and \textbf{1895}), but \textbf{1895} has a larger $\Delta E_{\mathrm{ST}}$ than ADBNA-Me-Mes and \textbf{845} has very small $f_{\mathrm{S}_1}$. There are a further four molecules with predicted emission above 600~nm, all with smaller $\Delta E_{\mathrm{ST}}$ than ADBNA-Me-Mes, with \textbf{1874} being the most promising with $f_{\mathrm{S}_1}=0.07$. At the violet end there is one outlier with predicted emission of 401~nm (\textbf{1585}), which has a smaller $\Delta E_{\mathrm{ST}}$ than the parent and $f_{\mathrm{S}_1}=0.02$, and various other molecules with predicted violet/deep blue emission. Seven molecules have $\Delta E_{\mathrm{ST}}<0.1$~eV, with blue to orange emission, and 16 have $0.1\leq\Delta E_{\mathrm{ST}}<0.2$~eV. Three molecules have both $\Delta E_{\mathrm{ST}}<0.3$~eV and $f_{\mathrm{S}_1}>0.1$ (\textbf{489}, \textbf{1332} and \textbf{1553}), which are blue or green. A total of 38 molecules have smaller $\Delta E_{\mathrm{ST}}$ and higher $f_{\mathrm{S}_1}$ than ADBNA-Me-Mes, with blue or green emission and a wide range of Tanimoto similarities with respect to the parent molecule. A range of heteroatoms are present in the molecules with small $\Delta E_{\mathrm{ST}}$. 

\paragraph{$\nu$-DABNA.}

Most child molecules are predicted to have blue emission like $\nu$-DABNA, with many also being green. There is one orange outlier (\textbf{1820}), with a similar $\Delta E_{\mathrm{ST}}$ to $\nu$-DABNA, but a much smaller $f_{\mathrm{S}_1}$, and two yellow emitters (\textbf{619} and \textbf{920}), with small $\Delta E_{\mathrm{ST}}$ (0.12/0.21 eV) and moderate $f_{\mathrm{S}_1}$ (0.02/0.07). There is also a violet outlier (\textbf{1499}), which is much smaller than the parent molecule and has a larger $\Delta E_{\mathrm{ST}}$ and smaller $f_{\mathrm{S}_1}$. Five molecules have $\Delta E_{\mathrm{ST}}<0.1$~eV, all of which are predicted to be green, with a range of $f_{\mathrm{S}_1}$. Three of these contain a P heteroatom, although other molecules containing P have $\Delta E_{\mathrm{ST}}$ values similar to or worse than $\nu$-DABNA, depending on the substitution position. Eight have $0.1\leq\Delta E_{\mathrm{ST}}<0.2$~eV, with $f_{\mathrm{S}_1}$ values up to 0.06, and a range of emission colours, from cyan to yellow. There are 13 molecules with $\Delta E_{\mathrm{ST}}<0.3$~eV and $f_{\mathrm{S}_1}>0.1$, mostly blue/cyan. Ten molecules have smaller $\Delta E_{\mathrm{ST}}$ and higher $f_{\mathrm{S}_1}$ than the parent molecule, all with blue emission and all but one with Tanimoto similarity $>0.77$ with respect to $\nu$-DABNA.

\paragraph{CzBN.}

Most child molecules are blue or green, in line with the blue emission of CzBN. There are various orange emitters, and two predicted red emitters (\textbf{17} and \textbf{497}), both with smaller $E_{\mathrm{ST}}$ but also smaller $f_{\mathrm{S}_1}$ than CzBN. There are eight violet emitters with predicted emission wavelengths below 430~nm, of which only \textbf{503} and \textbf{1197} have smaller $\Delta E_{\mathrm{ST}}$ than the parent, with \textbf{1197} having the larger $f_{\mathrm{S}_1}$ of the two (0.1 \emph{vs.}\ 0.03). Nineteen molecules have $\Delta E_{\mathrm{ST}}<0.2$~eV, of which six $\Delta E_{\mathrm{ST}}<0.1$~eV, which collectively span blue to red. A number of these contain either P or S heteroatoms. Only two molecules have both $\Delta E_{\mathrm{ST}}<0.3$~eV and $f_{\mathrm{S}_1}>0.1$: \textbf{629} and \textbf{1686}, which are blue and green respectively. A single molecule has both smaller $\Delta E_{\mathrm{ST}}$ and higher $f_{\mathrm{S}_1}$ than CzBN: \textbf{273}, which has almost identical predicted emission colour to CzBN, with the only difference in structure being the substitution of one of the H atoms with Cl.

\paragraph{2PTZBN.}

Most child molecules are blue or green, in line with the predicted emission of 2PTZBN, which is slightly blue-shifted compared to the green of experiment.  There are two red molecules (\textbf{977} and \textbf{1425}), as well as one orange (\textbf{1659}) and various other yellow/yellow-orange molecules. Of these both \textbf{977} and \textbf{1659} have a smaller $\Delta E_{\mathrm{ST}}$ than PTZBN, but \textbf{977} has a very small $f_{\mathrm{S}_1}$: 0.006 compared to 0.02 for \textbf{1659}. There are two violet outliers with predicted emission wavelengths below 420~nm (\textbf{447} and \textbf{697}), where \textbf{447} has a smaller $\Delta E_{\mathrm{ST}}$ than PTZBN, but also a smaller $f_{\mathrm{S}_1}$ (0.02). Sixteen molecules have $\Delta E_{\mathrm{ST}}<0.1$~eV, and 12 have $0.1\leq\Delta E_{\mathrm{ST}}<0.2$~eV, collectively spanning red to blue emission. Two molecules have both $\Delta E_{\mathrm{ST}}<0.3$~eV and $f_{\mathrm{S}_1}>0.1$: \textbf{1307} and \textbf{1852}. There are five molecules with both $\Delta E_{\mathrm{ST}}$ and higher $f_{\mathrm{S}_1}$ than 2PTZBN, all of which have high Tanimoto similarities with respect to 2PTZBN ($>0.7$) and very similar predicted emission wavelengths. Both I and P heteroatoms feature among these molecules, with other heteroatoms such as Cl appearing among the molecules with small $\Delta E_{\mathrm{ST}}$.

\paragraph{BN3.}

Most child molecules are yellow or green, in line with BN3, whose predicted emission is a yellowish green, with a wavelength that is slightly underestimated compared to experiment. Four molecules have predicted orange emission, of which all three have $\Delta E_{\mathrm{ST}}<0.3$~eV (\textbf{648}, \textbf{654} and \textbf{1496}), while $f_{\mathrm{S}_1}$ values range from 0.005 to 0.08. There are seven blue/violet emitters, ranging from cyan to violet, of which \textbf{1379} and \textbf{1738} are the only ones with $\Delta E_{\mathrm{ST}}<0.3$~eV, with $f_{\mathrm{S}_1}$ values of 0.003 and 0.04 respectively. Four of these, including \textbf{1379}, are significantly smaller than the parent molecule. Four molecules have $\Delta E_{\mathrm{ST}}<0.1$~eV,  with $f_{\mathrm{S}_1}$ ranging from $4\times 10^{-4}$ to 0.02; all with yellow predicted emission. Five have $0.1\leq\Delta E_{\mathrm{ST}}<0.2$~eV, with $f_{\mathrm{S}_1}$ ranging from 0.01 to 0.09, and again mostly yellow emission. There are 32 molecules with $\Delta E_{\mathrm{ST}}<0.3$~eV and $f_{\mathrm{S}_1}>0.1$, ranging from green to yellow emission. There are seven molecules with  smaller $\Delta E_{\mathrm{ST}}$ and larger $f_{\mathrm{S}_1}$ than BN3, all with a Tanimoto similarity above 0.8 with respect to BN3, and predicted green emission. 

\paragraph{2F-BN.}

The predicted emission wavelength for 2F-BN is slightly underestimated, being a bright blue rather than green. Most child molecules are therefore also predicted to be blue, with many also being green. There is one orange outlier (\textbf{234}) and one clear yellow outlier (\textbf{1971}), with \textbf{1971} having smaller $\Delta E_{\mathrm{ST}}$ than 2F-BN (0.31~eV), and $f_{\mathrm{S}_1}=0.08$. There are also some green-yellow emitters. There are two violet outliers (\textbf{743} and \textbf{1521}), which both have larger $\Delta E_{\mathrm{ST}}$ than 2F-BN.
Six molecules have $\Delta E_{\mathrm{ST}}<0.1$~eV, with $f_{\mathrm{S}_1}$ ranging from $6\times 10^{-4}$ to 0.02, and blue or green emission. Nine have $0.1\leq\Delta E_{\mathrm{ST}}<0.2$~eV, with $f_{\mathrm{S}_1}$ ranging from  0.001 to 0.13, all of which are predicted to be green, and two of which feature S heteroatoms. Overall, five molecules have $\Delta E_{\mathrm{ST}}<0.3$~eV and $f_{\mathrm{S}_1}>0.1$, all of which are blue or green.  There are eleven molecules with  smaller $\Delta E_{\mathrm{ST}}$ and larger $f_{\mathrm{S}_1}$ than 2F-BN, all with predicted blue emission and a high Tanimoto similarity with respect to the parent molecule ($>0.74$).

\paragraph{DBNO.}

Most child molecules are predicted to have green emission like DBNO, but there is a noticeable outlier in the red emission of \textbf{1290}. However it has a very small $f_{\mathrm{S}_1}$ (0.007), and slightly higher $\Delta E_{\mathrm{ST}}$ than DBNO. The next most red-shifted child molecule is \textbf{578}, which is predicted to emit yellow/yellow-green light, and has a higher $f_{\mathrm{S}_1}$ (0.33) and smaller $\Delta E_{\mathrm{ST}}$ compared to DBNO (0.27~eV). Three molecules are violet (\textbf{68}, \textbf{947} and \textbf{1902}); all have reasonable $f_{\mathrm{S}_1}$ ($>0.2$), but higher $\Delta E_{\mathrm{ST}}$ than DBNO. In common with \textbf{1290}, they are much smaller molecules than DBNO. Two molecules (\textbf{216} and \textbf{1168}) have $\Delta E_{\mathrm{ST}}<0.1$~eV, but have correspondingly small $f_{\mathrm{S}_1}$ (0.01 and 0.002). Two molecules (\textbf{492} and \textbf{1663}) have $0.1\leq\Delta E_{\mathrm{ST}}<0.2$~eV, with larger $f_{\mathrm{S}_1}$ (0.07 and 0.11); all four of these molecules are predicted to have green emission. Overall, four molecules have $\Delta E_{\mathrm{ST}}<0.3$~eV and $f_{\mathrm{S}_1}>0.1$; two are blue and two green. There are five emitters with smaller $\Delta E_{\mathrm{ST}}$ and larger $f_{\mathrm{S}_1}$ than DBNO, all green and with high Tanimoto similarity with respect to the parent  ($>0.88$).

\section{Conclusions}

In this work we presented an HTVS workflow for identifying potential new TADF emitters, using the STONED algorithm to generate an initial library of child molecules based on random mutations to a set of parent molecules. Subsequent steps included initial structural and synthesisability filters, as well as quantum chemistry calculations. Benchmark calculations were performed to justify the use of intermediate filters based on the HOMO-LUMO gap and spatial overlap, leading to significant computational savings due to eliminating unsuitable molecules before the more expensive DFT geometry optimisations and TDDFT calculations. The STONED algorithm allows for a balance between exploring relevant chemical space and achieving diversity, but it proved to be ineffective at rediscovering existing TADF emitters. Nonetheless, the workflow was able to identify a number of molecules worthy of further investigation, without any bias towards existing design rules. In future, it might be interesting to combine random mutations with a more targeted approach, e.g.\ guiding mutations toward molecular frameworks with established TADF characteristics or integrating domain-specific design principles into the workflow.

The workflow was applied to twenty known TADF parent emitters, split between D-A and MR emitters. While the number of molecules passing all filters varied significantly across parent molecules, in each case a number of potentially promising emitters were identified, covering a range of predicted emission colours. In total 171 D-A and 111 MR molecules were predicted to have both smaller $\Delta E_{\mathrm{ST}}$ and larger $f_{\mathrm{S}_1}$ than their respective parent molecules. 
The data produced in this work also offers the possibility of being used to explore relationships between atomic structure and target properties which go beyond known design rules, potentially also supplemented by further runs with additional parent molecules, to generate a larger dataset.

The current work focused on using robust approaches with relatively low computational cost, in order to screen large numbers of molecules. In order to predict $\Delta E_{\mathrm{ST}}$, $f_{\mathrm{S}_1}$ and emission wavelengths more accurately and thus determine which might be worth exploring experimentally, it would be desirable to perform further calculations in future using a higher level of theory for the most promising candidate molecules. This would include for example using tuned range-separated hybrid functionals for the TDDFT calculations of D-A emitters, alternative methods beyond TDDFT for the MR emitters, and excited state geometry optimisations for both classes.





\section*{Data availability}

Data for this article, including relaxed molecular geometries and calculated properties, are available at Gitlab at \url{https://gitlab.com/lratcliff/tadf-htvs}.

\section*{Acknowledgements}

LER acknowledges support from an EPSRC Early Career Research Fellowship (EP/P033253/1). KT acknowledges support from the Engineering and Physical
Sciences Research Council (EP/S515085/1). We are also grateful for computational support from the UK national high performance computing service, ARCHER2, for which access was obtained via the UKCP consortium and funded by EPSRC grant ref EP/X035891/1.  Parts of this work were also carried out using the computational facilities of the Advanced Computing Research Centre, University of Bristol -- http://www.bristol.ac.uk/acrc/. The authors would like to thank Louis Beal, William Dawson and Luigi Genovese, for their technical input, including advice on the remote submission workflows.

\bibliography{refs} 
\bibliographystyle{rsc} 
\end{document}


\title{{\LARGE Supplementary Information}\\A High Throughput Virtual Screening Approach for Identifying Thermally Activated Delayed Fluorescence-Based Emitters}

\author{Kritam Thapa}  
\email{These authors contributed equally to this work}
\affiliation{\ICL} 
\author{Jennifer I.\ Jones}     
\email{These authors contributed equally to this work}
\affiliation{\Bristol} 
\author{Laura E.\ Ratcliff} 
\email{laura.ratcliff@bristol.ac.uk}
\affiliation{\Bristol}
\affiliation{\UiT}

\date{\today}

\maketitle

\begingroup
\renewcommand\thefootnote{\dagger}
\footnotetext{These authors contributed equally to this work.}
\endgroup

\begin{table*}[]
\centering
\begin{threeparttable}
\caption{The employed SMILES strings for the different parent molecules.}
\label{tab:parent_smiles}
\begin{tabular*} {1.0\textwidth}{l @{\extracolsep{\fill}} l}
\hline \hline
Name & SMILES\\
\cline{1-1}\cline{2-2}\\[-2.5ex]
D-A\\
2CzPN & N\#Cc1cc(-n2c3ccccc3c3ccccc32)c(-n2c3ccccc3c3ccccc32)cc1C\#N \\
4CzIPN & N\#Cc1c(-n2c3ccccc3c3ccccc32)c(C\#N)c(-n2c3ccccc3c3ccccc32)c(-n2c3ccccc3c3ccccc32)c1-n1c2ccccc2c2 \\
 & ccccc21 \\
TXO-TPA & O=C5c1ccccc1S(=O)(=O)c6ccc(c4ccc(N(c2ccccc2)c3ccccc3)cc4)cc56 \\
CCO-1 & O=c8c7ccc(N6c1ccccc1C4(c2ccccc2Oc3ccccc34)c5ccccc56)cc7oc9c8ccc\%11c9c\%10ccccc\%10n\%11c\%\\
 & 12ccccc\%12 \\
Pm2 & CC5(C)c1ccccc1N(c3ccc(c2cnc(C\#N)nc2)cc3)c4ccccc45 \\
3DMAC-BP-CN & CC1(C)c2ccccc2N(c2ccc3c(c2)c2cc(N4c5ccccc5C(C)(C)c5ccccc54)ccc2c2nc4cc(N5c6ccccc6C(C)(C) \\
 & c6ccccc65)\\
& c(C\#N)cc4nc32)c2ccccc21 \\
PXZ-TRZ & c7ccc(c6nc(c1ccccc1)nc(c5ccc(n4c2ccccc2oc3ccccc34)cc5)n6)cc7 \\
DHPZ-2TRZ & c1ccc(-c2nc(-c3ccccc3)nc(-c3ccc(N4c5ccccc5N(c5ccc(-c6nc(-c7ccccc7)nc(-c7ccccc7)n6)cc5)c5ccccc54) \\
 & cc3)n2)cc1 \\
SpiroAC-TRZ & c\%10ccc(c9nc(c1ccccc1)nc(c8ccc(N7c2ccccc2C5(c3ccccc3c4ccccc45)c6ccccc67)cc8)n9)cc\%10 \\
ACRFLCN & N\#Cc1ccc2c(c1)C1(c3cc(C\#N)ccc3-2)c2ccccc2N(c2ccccc2)c2ccccc21 \\
\cline{1-1}\cline{2-2}\\[-2.5ex]
MR\\
DOBNA & c3ccc2B4c1ccccc1Oc5cccc(Oc2c3)c45 \\
3Cz-DiKTa & O=c\%12c4cc(n3c1ccccc1c2ccccc23)ccc4n\%14c8ccc(n7c5ccccc5c6ccccc67)cc8c(=O)c\%13cc(n\%11c9ccccc9c\%\\
 & 10ccccc\%10\%11)cc\%12c\%13\%14 \\
DABNA-1 & b12c3c(n(c4ccccc14)c1ccccc1)cccc3n(c1c2cccc1)c1ccccc1 \\
ADBNA-Me-Mes & Cc1ccc5c(c1)B(c2c(C)cc(C)cc2C)c6cc(C)cc7B(c3c(C)cc(C)cc3C)c4cc(C)ccc4N5c67 \\
$\nu$-DABNA & c1ccc(N(c2ccccc2)c2cc3c4c(c2)N(c2ccccc2)c2cc5c(cc2B4c2ccccc2N3c2ccccc2)B2c3ccccc3N(c3ccccc3)c3cc\\
 & (N(c4ccccc4)c4ccccc4)cc(c32)N5c2ccccc2)cc1 \\
CzBN & c7cc2c6B(c3cccc4c1ccccc1n2c34)c8cccc9c5ccccc5n(c6c7)c89 \\
2PTZBN & c7cc2c6B(c3cccc4sc1ccccc1n2c34)c8cccc9sc5ccccc5n(c6c7)c89 \\
BN3 & c\%17ccc(N(c1ccccc1)c2ccc\%12c(c2)c\%14cc(N(c3ccccc3)c4ccccc4)cc\%13B\%10c\%15cc(N(c5ccccc5)c6ccccc6)\\
 & cc\%16c9cc(N(c7ccccc7)c8ccccc8)ccc9n(c\%11cccc(c\%10\%11)n\%12c\%13\%14)c\%15\%16)cc\%17 \\
2F-BN & CC(C)(C)c1ccc6c(c1)c8cc(C(C)(C)C)cc7B4c9cc(C(C)(C)C)cc\%10c2cc(C(C)(C)C)ccc2n(c5cc(c3ccccc3F)\\
 & cc(c45)n6c78)c9\%10 \\
DBNO & CC(C)(C)c1ccc7c(c1)c9cc(C(C)(C)C)cc8B5c4cc3Oc2cccc\%11c2B(c3cc4Oc6cccc(c56)n7c89)c\%12cc\\
 & (C(C)(C)C)cc\%13c\%10cc(C(C)(C)C)ccc\%10n\%11c\%12\%13 \\
 \hline \hline
\end{tabular*}
\end{threeparttable}
\end{table*}

\begin{table*}[]
\centering
\begin{threeparttable}
\caption{A summary of key calculated and corresponding experimental properties for the parent molecules. Theoretical values for $\Delta E_{\mathrm{ST}}$, $f_{\mathrm{S}_1}$ and S$_1$ come directly from vertical TDDFT calculations, while the theoretical emission wavelength is predicted from the S$_1$ energy using the linear fit approach described in the main text. Experimental references are given below: where possible values measured in toluene were employed to maximise consistency, but some measurements were taken in another solvent or in various thin films.}
\label{tab:parent_molecules}
\begin{tabular*} {1.0\textwidth}{l @{\extracolsep{\fill}} llllllll}
\hline \hline
 & Number & & \multicolumn{2}{c}{$\Delta E_{\mathrm{ST}}$ / eV} & $f_{\mathrm{S}_1}$  & S$_1$ / eV & \multicolumn{2}{c}{Emission Wavelength / nm}\\
 \cline{4-5}\cline{6-6}\cline{7-7}\cline{8-9}\\[-2.5ex]
Name & of Atoms & SAscore & Theor. & Exp. & Theor. & Theor. & Theor. & Exp.\\
\cline{1-1}\cline{2-2}\cline{3-3}\cline{4-4}\cline{5-5}\cline{6-6}\cline{7-7}\cline{8-8}\cline{9-9}\\[-2.5ex]
D-A\\
2CzPN & 54 & 2.3 & 0.28 & 0.21\tnote{a} & 5.5E-02 & 2.79 & 488 & 473\tnote{a} \\
4CzIPN & 94 & 2.8 & 0.17 & 0.083\tnote{b} & 5.9E-02 & 2.59 & 510 & 507\tnote{b} \\
TXO-TPA & 57 & 2.4 & 0.10 & 0.052\tnote{c} & 4.5E-02 & 2.23 & 549 & 586\tnote{c} \\
CCO-1 & 85 & 3.6 & 0.01 & 0.063\tnote{d} & 8.7E-06 & 2.70 & 497 & 465\tnote{d} \\
Pm2 & 50 & 2.4 & 0.0043 & 0.09\tnote{e} & 2.0E-05 & 2.24 & 548 & 524\tnote{e} \\
3DMAC-BP-CN & 122 & 3.6 & 0.014 & 0.02\tnote{f} & 3.5E-04 & 1.51 & 629 & 619\tnote{f} \\
PXZ-TRZ & 60 & 2.2 & 0.12 & 0.06\tnote{g} & 1.2E-01 & 2.25 & 547 & 545\tnote{h} \\
DHPZ-2TRZ & 98 & 2.4 & 0.0046 & nearly zero\tnote{i} & 6.2E-08 & 1.67 & 612 & 650\tnote{i} \\
SpiroAC-TRZ & 80 & 2.9 & 0.0063 & 0.072\tnote{j} & 3.5E-05 & 2.42 & 528 & 479\tnote{j} \\
ACRFLCN & 55 & 2.9 & 0.011 & 0.028\tnote{k} & 4.3E-04 & 2.51 & 519 & 486\tnote{g} \\
\cline{1-1}\cline{2-2}\cline{3-3}\cline{4-4}\cline{5-5}\cline{6-6}\cline{7-7}\cline{8-8}\cline{9-9}\\[-2.5ex]
MR\\
DOBNA & 32 & 3.0 & 0.58 & 0.18\tnote{l} & 1.5E-01 & 3.52 & 407 & 398\tnote{l} \\
3Cz-DiKTa & 94 & 2.8 & 0.30 & 0.11\tnote{m} & 6.0E-02 & 2.29 & 543 & 534\tnote{m} \\
DABNA-1 & 54 & 2.5 & 0.55 & 0.2\tnote{n} & 2.5E-01 & 3.21 & 441 & 462\tnote{n} \\
ADBNA-Me-Mes & 81 & 3.5 & 0.57 & 0.2\tnote{o} & 3.0E-01 & 3.01 & 462 & 484\tnote{o} \\
$\nu$-DABNA & 140 & 4.0 & 0.34 & 0.017\tnote{p} & 8.0E-01 & 2.93 & 472 & 468\tnote{p} \\
CzBN & 50 & 2.9 & 0.46 & 0.12\tnote{q} & 4.5E-01 & 2.94 & 470 & 479\tnote{q} \\
2PTZBN & 52 & 3.2 & 0.45 & 0.15\tnote{r} & 2.7E-01 & 2.82 & 484 & 510\tnote{r} \\
BN3 & 138 & 3.9 & 0.31 & 0.09\tnote{s} & 4.0E-01 & 2.27 & 545 & 562\tnote{s} \\
2F-BN & 108 & 3.7 & 0.45 & 0.16\tnote{t} & 4.9E-01 & 2.86 & 480 & 501\tnote{t} \\
DBNO & 118 & 4.3 & 0.39 & 0.17\tnote{u} & 6.9E-01 & 2.67 & 501 & 500\tnote{u} \\

 \hline \hline
\end{tabular*}
  \begin{tablenotes}
    \item[a] Ref.~\onlinecite{Kim2017}
    \item[b] Ref.~\onlinecite{Uoyama2012}
    \item[c] Ref.~\onlinecite{Wang2014}
    \item[d] Ref.~\onlinecite{Fu2023}
    \item[e] Ref.~\onlinecite{Pan2016}
    \item[f] Ref.~\onlinecite{Huang2021}
    \item[g] Ref.~\onlinecite{Huang2013}
    \item[h] Ref.~\onlinecite{Tanaka2012}
    \item[i] Ref.~\onlinecite{Lee2015}
    \item[j] Ref.~\onlinecite{Lin2016}
    \item[k] Ref.~\onlinecite{Mehes2012}
    \item[l] Ref.~\onlinecite{Ikeda2020}
    \item[m] Ref.~\onlinecite{Wu2022}
    \item[n] Ref.~\onlinecite{Hatakeyama2016}
    \item[o] Ref.~\onlinecite{Oda2019}
    \item[p] Ref.~\onlinecite{Kondo2019}
    \item[q] Ref.~\onlinecite{Xu2021}
    \item[r] Ref.~\onlinecite{Hua2021}
    \item[s] Ref.~\onlinecite{Qi2021}
    \item[t] Ref.~\onlinecite{Cai2022}
    \item[u] Ref.~\onlinecite{Qi2021}
  \end{tablenotes}
\end{threeparttable}
\end{table*}

\begin{table}[ht]
\centering
\begin{threeparttable}
\caption{Thresholds employed in the workflow, including the experimental emission wavelength cut-off (experimental adiabatic $\Delta E_{\mathrm{ST}}$), the corresponding calculated vertical S$_1$ value (calculated vertical $\Delta E_{\mathrm{ST}}$), determined according to the fit in the S$_1$ ($\Delta E_{\mathrm{ST}}$) experimental benchmarking plots in the main paper. Also given is the resulting HOMO-LUMO gap (spatial overlap) threshold, determined using the results of the S$_1$ ($\Delta E_{\mathrm{ST}}$) benchmark plots generated for selected child molecules. Where relevant, separate thresholds are given for D-A and MR molecules.}
\label{tab:thresholds}
\begin{tabular*} {0.65\textwidth}{l @{\extracolsep{\fill}} rr}
\hline \hline
&  D-A & MR \\
\cline{1-1}\cline{2-2}\cline{3-3}\\[-2.5ex]
Experimental Emission Wavelength / nm & 380 & 380	\\
Calculated Vertical S$_1$	/ eV & 3.759 & 3.759 \\
HOMO-LUMO Gap / eV	& 2.941 & 2.941\\
\cline{1-1}\cline{2-2}\cline{3-3}\\[-2.5ex]
Experimental Adiabatic $\Delta E_{\mathrm{ST}}$ / eV & 0.210 & 0.210\\
Calculated Vertical $\Delta E_{\mathrm{ST}}$ / eV & 0.371	& 0.607\\
HOMO-LUMO Spatial Overlap	& 0.470 & 0.650\\
 \hline \hline
\end{tabular*}
\end{threeparttable}
\end{table}

\begin{table}[h]
\centering
\begin{threeparttable}
\caption{Number of molecules which were retained/eliminated due to being predicted to have a non-visible emission wavelength or a too large $\Delta E_{\mathrm{ST}}$, using the HOMO-LUMO gap and HOMO-LUMO spatial overlap thresholds respectively. Shown are both the number correctly retained/eliminated, i.e.\ for which the prediction was proven to be correct according to subsequent TDDFT calculations, and needlessly retained/incorrectly eliminated (lost), i.e.\ the prediction was proven to be \emph{incorrect} according to subsequent TDDFT calculations, leading to molecules being erroneously lost or needlessly calculated. Values are given for a subset of 200 molecules for each of the listed parent molecules (including the parent itself and 199 child molecules).}
\label{tab:benchmark_summary}
\begin{tabular*} {0.65\textwidth}{l @{\extracolsep{\fill}} rrrrrr}
\hline \hline
&  \multicolumn{2}{c}{Retained} & \multicolumn{2}{c}{Eliminated} \\
\cline{2-3}\cline{4-5}\\[-2.5ex]
&  Correctly & Needlessly & Correctly & Lost \\
\cline{1-1}\cline{2-2}\cline{3-3}\cline{4-4}\cline{5-5}\\[-2.5ex]
Emission Wavelength\\
D-A\\
2CzPN & 194 & 0 & 6 & 0 \\
TXO-TPA & 198 & 1 & 1 & 0 \\
PXZ-TRZ & 196 & 2 & 2 & 0 \\
DHPZ-2TRZ & 200 & 0 & 0 & 0 \\
ACRFLCN & 196 & 3 & 1 & 0 \\
MR\\
DOBNA & 179 & 10 & 9 & 2 \\
3Cz-DiKTa & 200 & 0 & 0 & 0 \\
DABNA-1 & 198 & 1 & 1 & 0 \\
CzBN & 200 & 0 & 0 & 0 \\
2PTZBN & 200 & 0 & 0 & 0 \\
\cline{1-1}\cline{2-2}\cline{3-3}\cline{4-4}\cline{5-5}\\[-2.5ex]
\cline{1-1}\cline{2-2}\cline{3-3}\cline{4-4}\cline{5-5}\\[-2.5ex]
$\Delta E_{\mathrm{ST}}$\\
D-A\\
2CzPN & 58 & 46 & 94 & 2 \\
TXO-TPA & 141 & 26 & 32 & 1 \\
PXZ-TRZ & 126 & 33 & 41 & 0 \\
DHPZ-2TRZ & 157 & 25 & 15 & 3 \\
ACRFLCN & 145 & 35 & 19 & 1 \\
MR\\
DOBNA & 81 & 73 & 44 & 2 \\
3Cz-DiKTa & 171 & 20 & 9 & 0 \\
DABNA-1 & 125 & 50 & 25 & 0 \\
CzBN & 124 & 45 & 28 & 3 \\
2PTZBN & 169 & 18 & 12 & 1 \\
 \hline \hline
\end{tabular*}
\end{threeparttable}
\end{table}

\clearpage
\begin{figure*}[t]
 \centering
 \begin{subfigure}[t]{0.48\textwidth}
    \centering
    \includegraphics[scale=0.5]{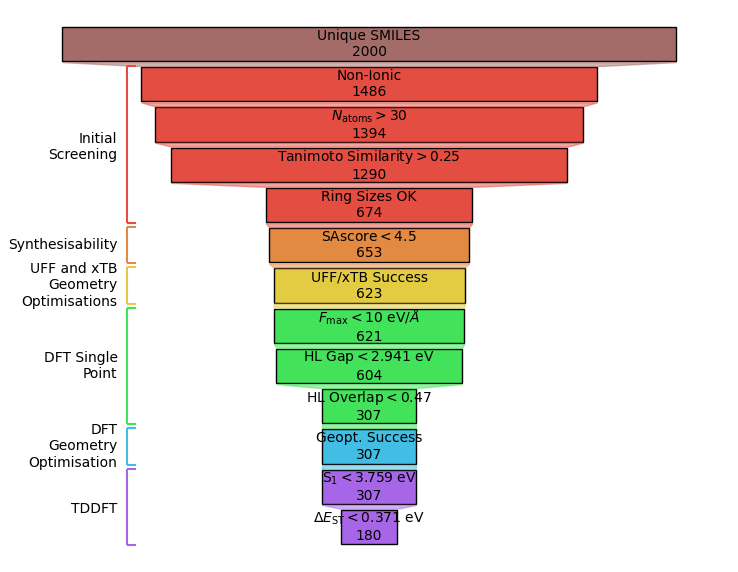}
    \caption{2CzPN}
\end{subfigure}
\hspace{2pt}
 \begin{subfigure}[t]{0.48\textwidth}
    \centering
    \includegraphics[scale=0.5]{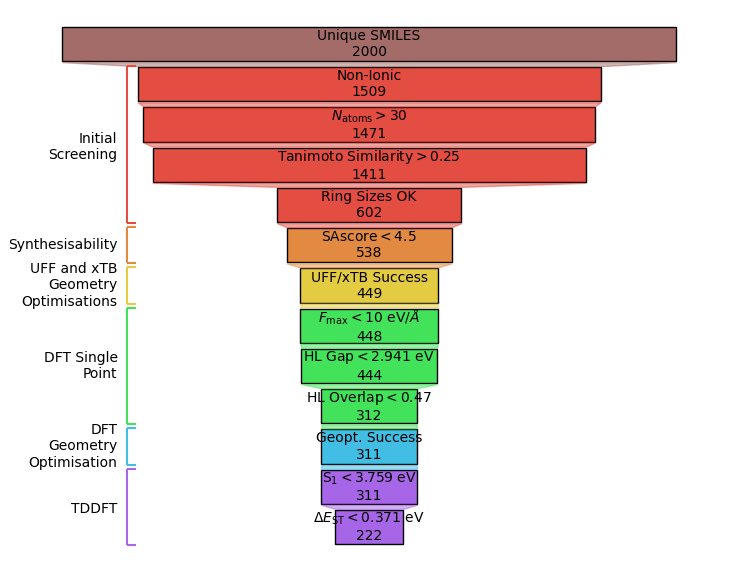}
    \caption{4CzIPN}
\end{subfigure}
 \begin{subfigure}[t]{0.48\textwidth}
    \centering
    \includegraphics[scale=0.5]{funnel_TXO-TPA.png}
    \caption{TXO-TPA}
\end{subfigure}
\hspace{2pt}
 \begin{subfigure}[t]{0.48\textwidth}
    \centering
    \includegraphics[scale=0.5]{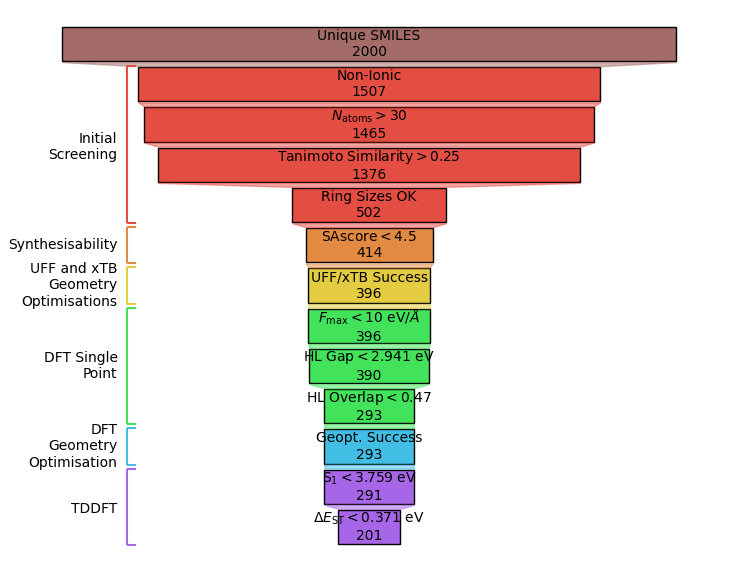}
    \caption{CCO-1}
\end{subfigure}
 \begin{subfigure}[t]{0.48\textwidth}
    \centering
    \includegraphics[scale=0.5]{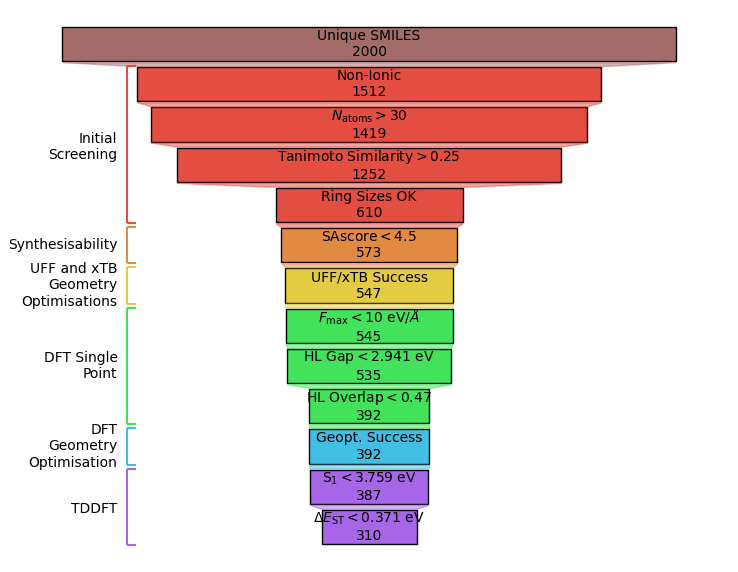}
    \caption{Pm2}
\end{subfigure}
\hspace{2pt}
 \begin{subfigure}[t]{0.48\textwidth}
    \centering
    \includegraphics[scale=0.5]{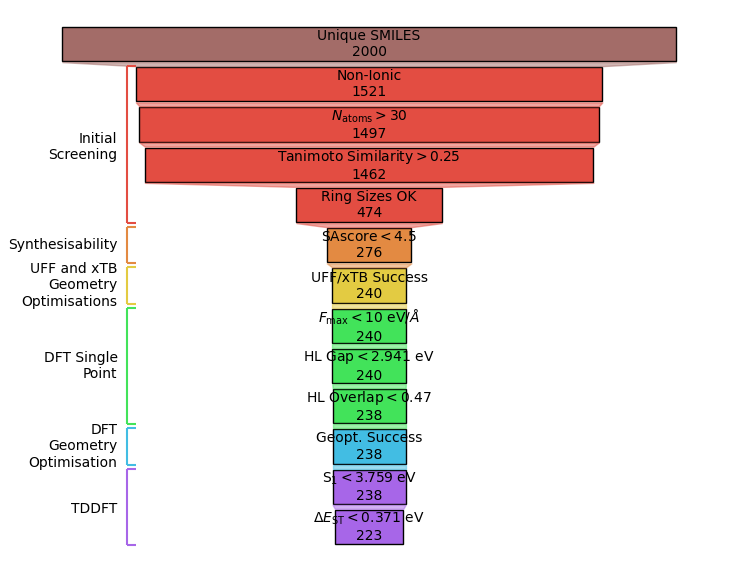}
    \caption{3DMAC-BP-CN}
\end{subfigure}
\hspace{2pt}
\caption{Funnel depiction of the various calculation steps for a subset of D-A parent molecules.}
\label{fig:si_funnels1}
\end{figure*} 

\begin{figure*}[t]
 \centering
 \begin{subfigure}[t]{0.48\textwidth}
    \centering
    \includegraphics[scale=0.5]{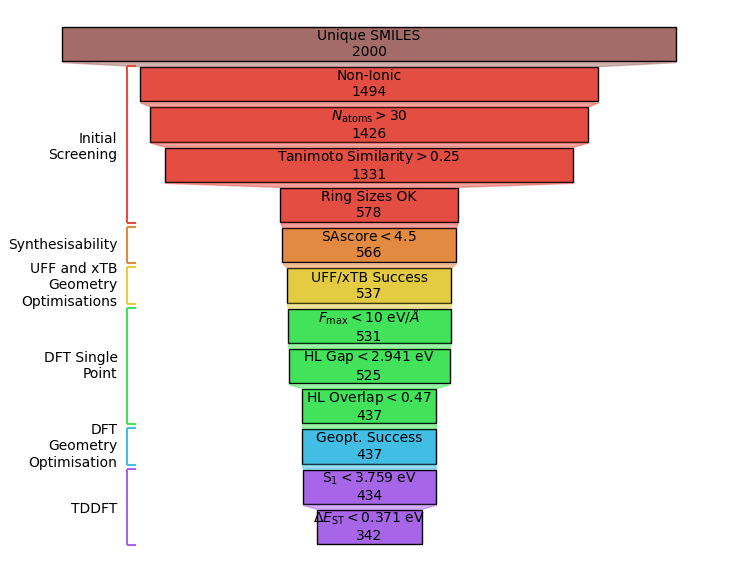}
    \caption{PXZ-TRZ}
\end{subfigure}
\hspace{2pt}
 \begin{subfigure}[t]{0.48\textwidth}
    \centering
    \includegraphics[scale=0.5]{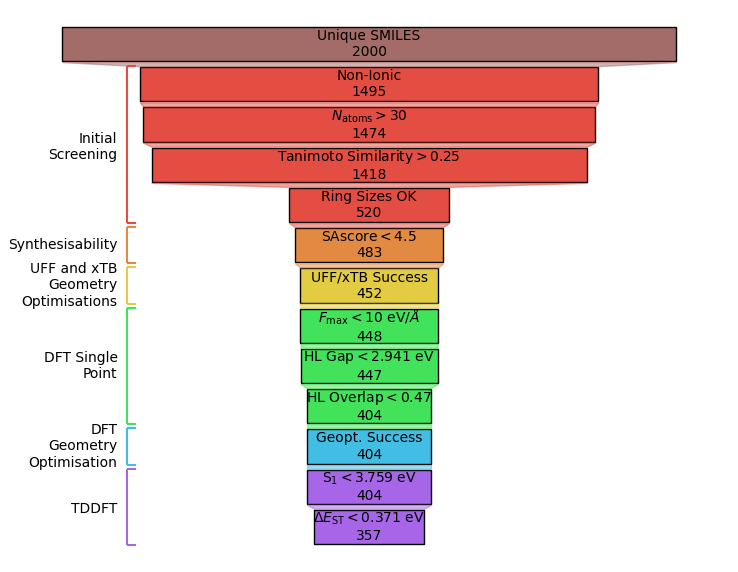}
    \caption{DHPZ-2TRZ}
\end{subfigure}
 \begin{subfigure}[t]{0.48\textwidth}
    \centering
    \includegraphics[scale=0.5]{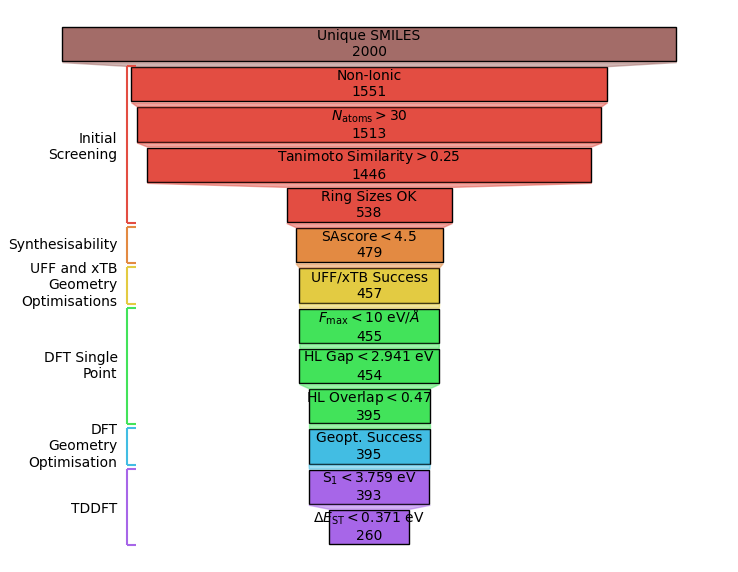}
    \caption{SpiroAC-TRZ}
\end{subfigure}
\hspace{2pt}
 \begin{subfigure}[t]{0.48\textwidth}
    \centering
    \includegraphics[scale=0.5]{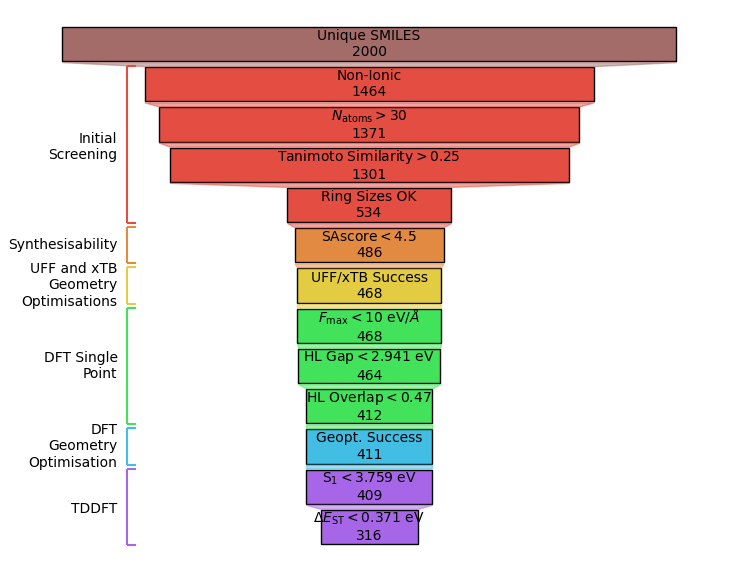}
    \caption{ACRFLCN}
\end{subfigure}
\hspace{2pt}
\caption{Funnel depiction of the various calculation steps for a subset of D-A parent molecules.}
\label{fig:si_funnels2}
\end{figure*}

\begin{figure*}[t]
 \centering
 \begin{subfigure}[t]{0.48\textwidth}
    \centering
    \includegraphics[scale=0.5]{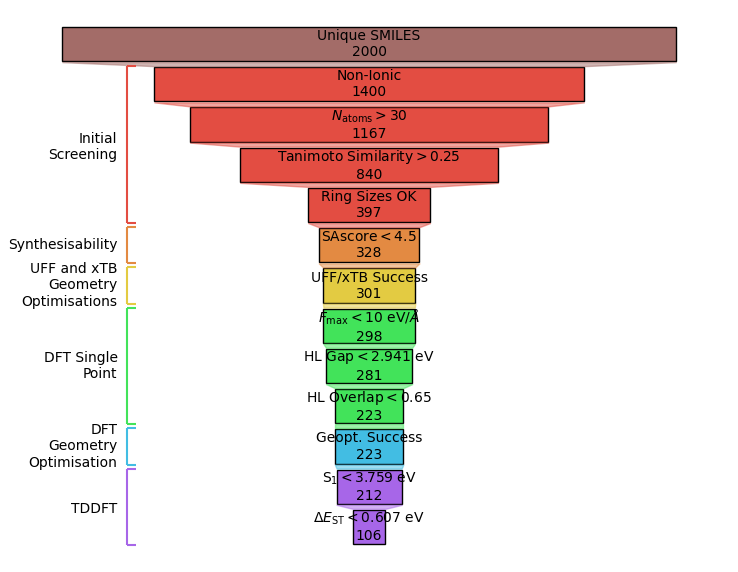}
    \caption{DOBNA}
\end{subfigure}
\hspace{2pt}
 \begin{subfigure}[t]{0.48\textwidth}
    \centering
    \includegraphics[scale=0.5]{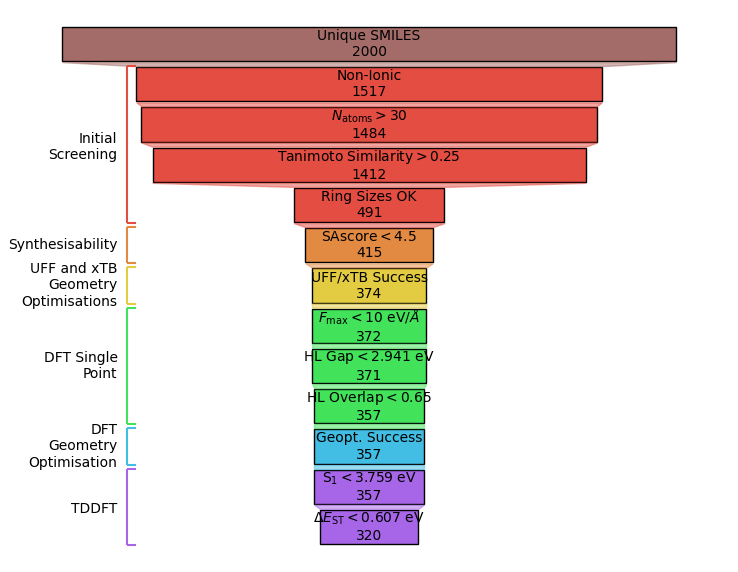}
    \caption{3Cz-DiKTa}
\end{subfigure}
 \begin{subfigure}[t]{0.48\textwidth}
    \centering
    \includegraphics[scale=0.5]{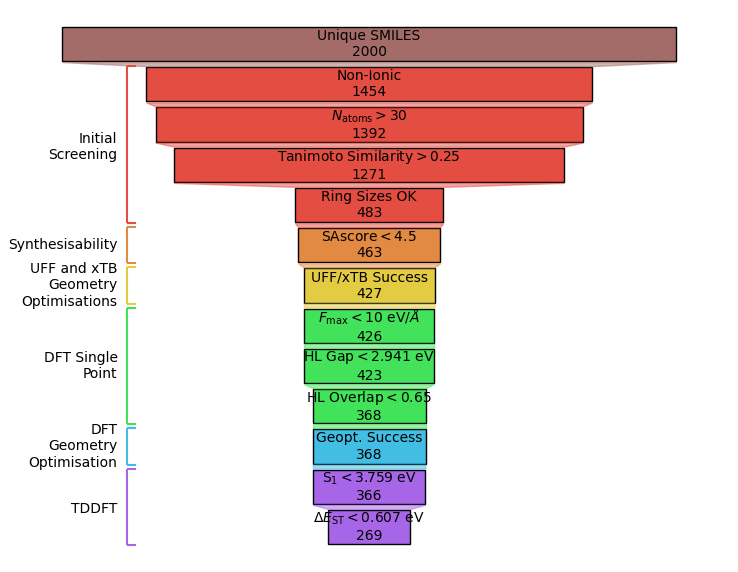}
    \caption{DABNA-1}
\end{subfigure}
\hspace{2pt}
 \begin{subfigure}[t]{0.48\textwidth}
    \centering
    \includegraphics[scale=0.5]{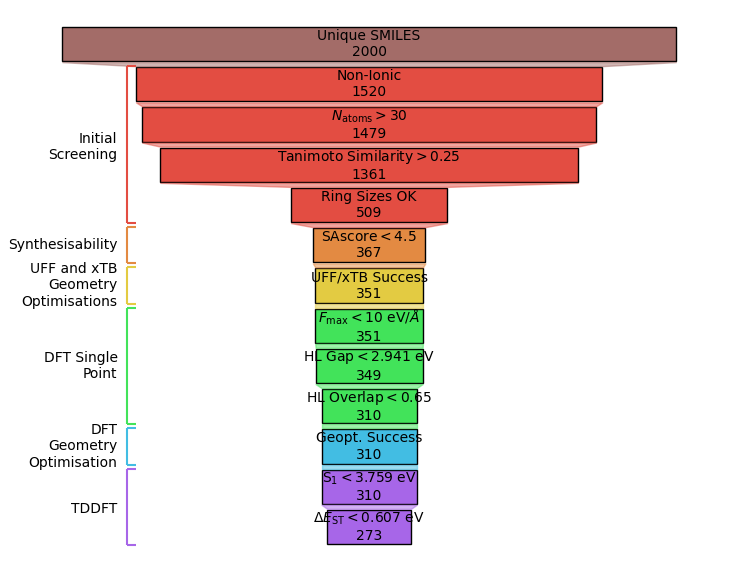}
    \caption{ADBNA-Me-Mes}
\end{subfigure}
 \begin{subfigure}[t]{0.48\textwidth}
    \centering
    \includegraphics[scale=0.5]{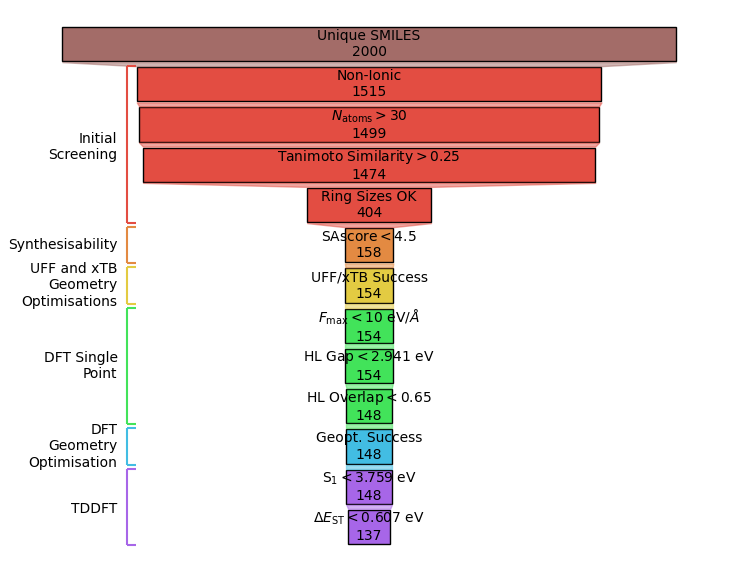}
    \caption{$\nu$-DABNA}
\end{subfigure}
\hspace{2pt}
 \begin{subfigure}[t]{0.48\textwidth}
    \centering
    \includegraphics[scale=0.5]{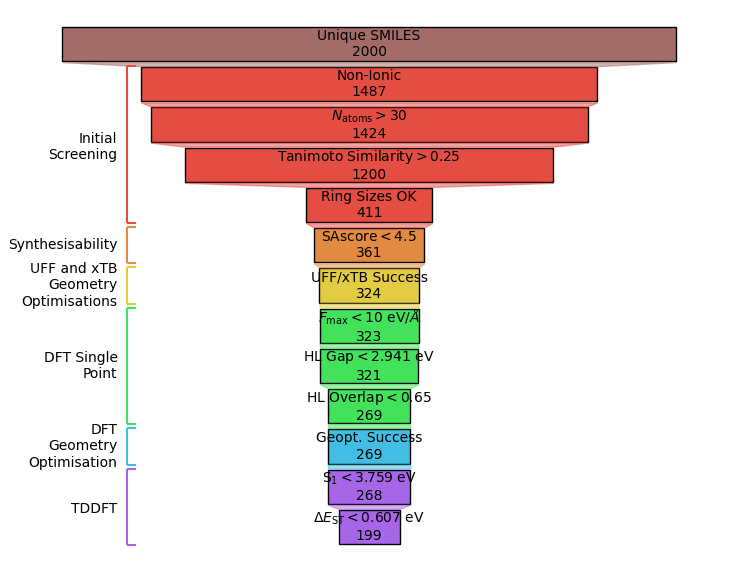}
    \caption{CzBN}
\end{subfigure}
\hspace{2pt}
\caption{Funnel depiction of the various calculation steps for a subset of MR parent molecules.}
\label{fig:si_funnels3}
\end{figure*}

\begin{figure*}[t]
 \centering
 \begin{subfigure}[t]{0.48\textwidth}
    \centering
    \includegraphics[scale=0.5]{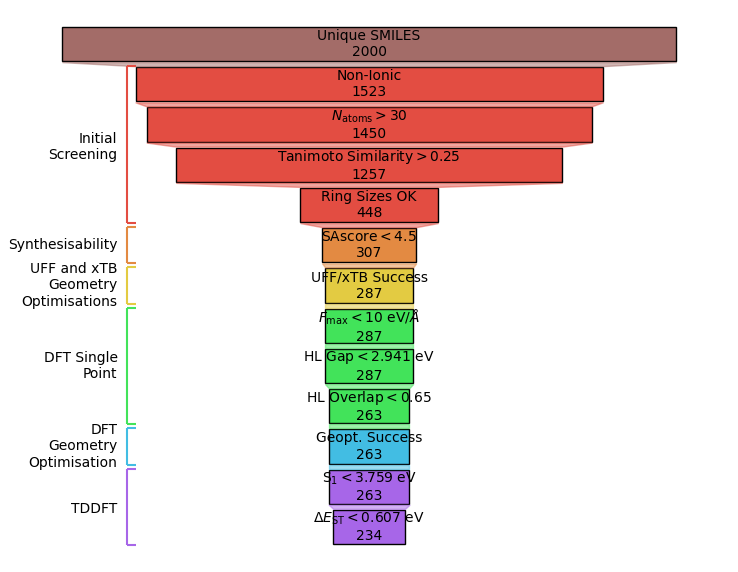}
    \caption{2PTZBN}
\end{subfigure}
\hspace{2pt}
 \begin{subfigure}[t]{0.48\textwidth}
    \centering
    \includegraphics[scale=0.5]{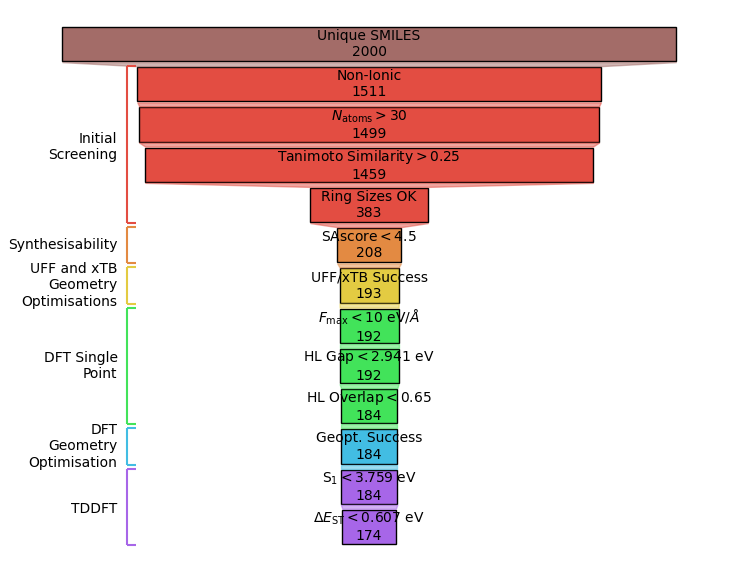}
    \caption{BN3} 
\end{subfigure}
 \begin{subfigure}[t]{0.48\textwidth}
    \centering
    \includegraphics[scale=0.5]{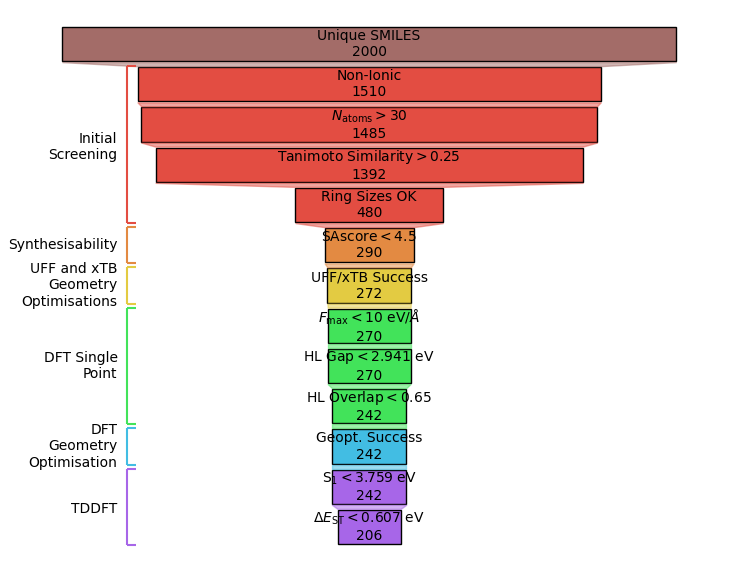}
    \caption{2F-BN}
\end{subfigure}
\hspace{2pt}
 \begin{subfigure}[t]{0.48\textwidth}
    \centering
    \includegraphics[scale=0.5]{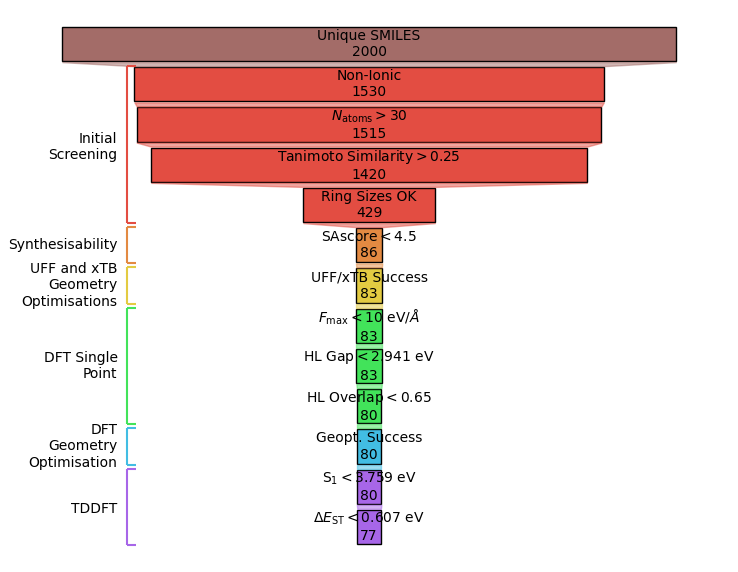}
    \caption{DBNO}
\end{subfigure}
\hspace{2pt}
\caption{Funnel depiction of the various calculation steps for a subset of MR parent molecules.}
\label{fig:si_funnels4}
\end{figure*}

 \begin{figure}[ht]
    \centering
    \includegraphics[scale=0.5]{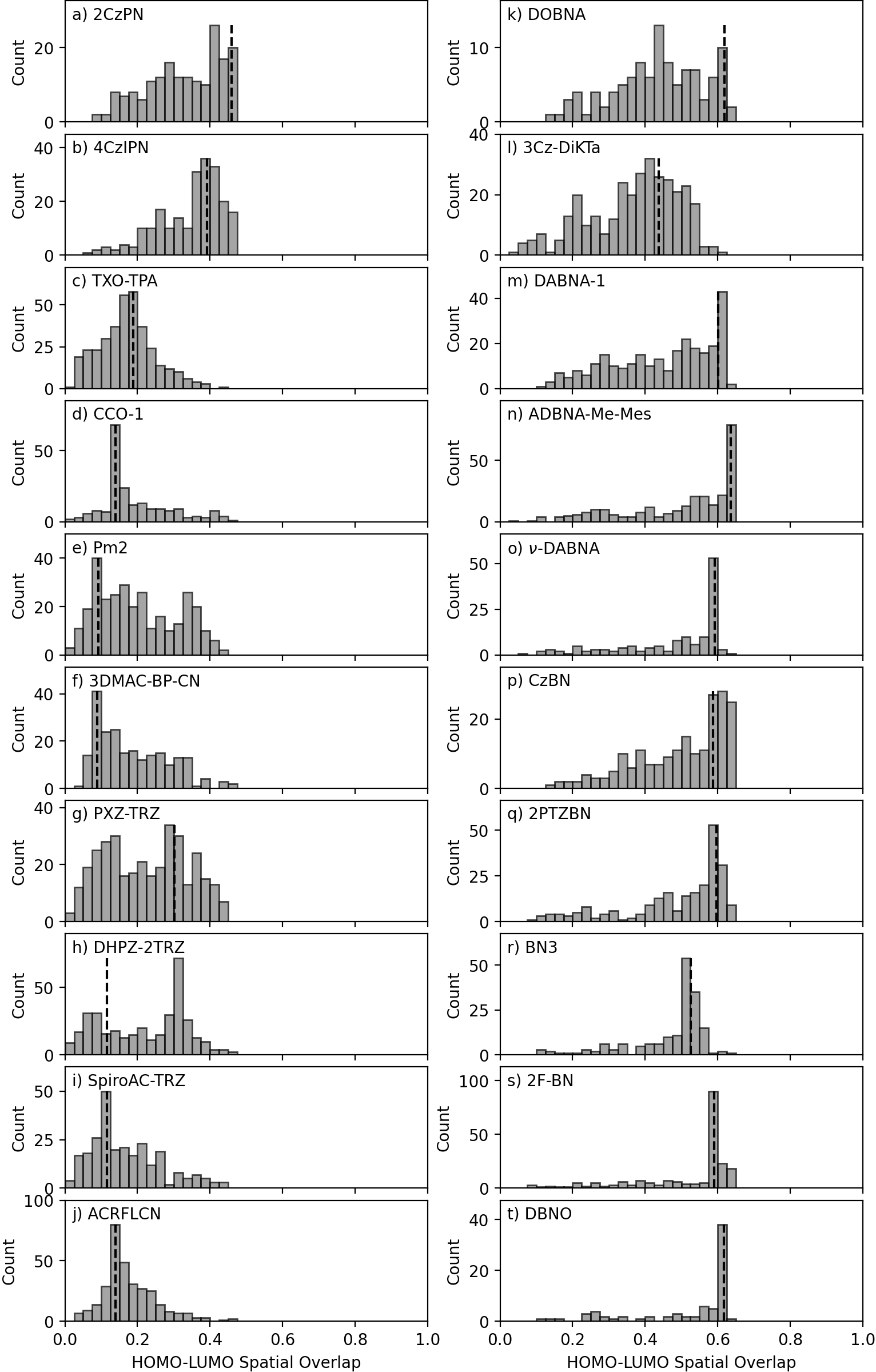}
    \caption{HOMO-LUMO spatial overlap values for molecules which passed all workflow filters. Values for parent molecules are depicted with a dashed vertical line.
    \label{fig:hl_overlap}}
\end{figure}

\begin{figure*}[t]
 \centering
 \begin{subfigure}[t]{0.48\textwidth}
    \centering
    \includegraphics[scale=0.5]{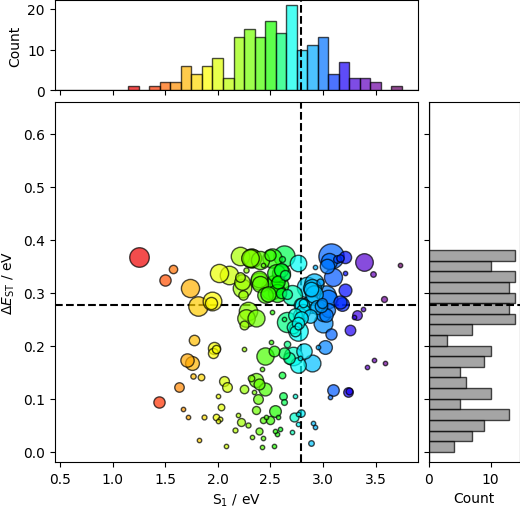}
    \caption{2CzPN}
\end{subfigure}
\hspace{2pt}
 \begin{subfigure}[t]{0.48\textwidth}
    \centering
    \includegraphics[scale=0.5]{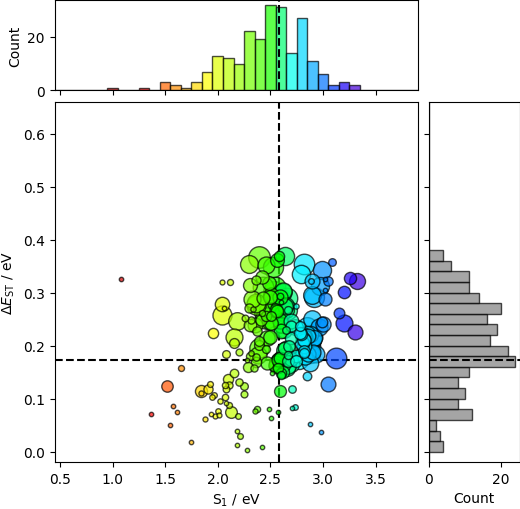}
    \caption{4CzIPN}
\end{subfigure}
 \begin{subfigure}[t]{0.48\textwidth}
    \centering
    \includegraphics[scale=0.5]{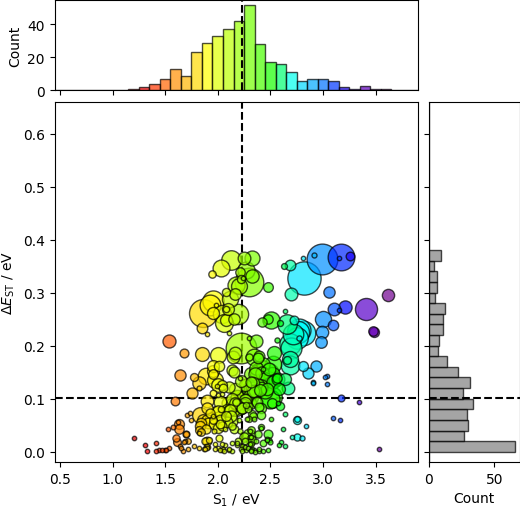}
    \caption{TXO-TPA}
\end{subfigure}
\hspace{2pt}
 \begin{subfigure}[t]{0.48\textwidth}
    \centering
    \includegraphics[scale=0.5]{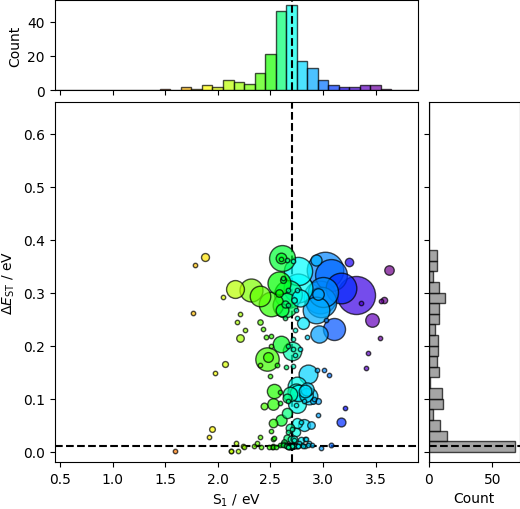}
    \caption{CCO-1}
\end{subfigure}
 \begin{subfigure}[t]{0.48\textwidth}
    \centering
    \includegraphics[scale=0.5]{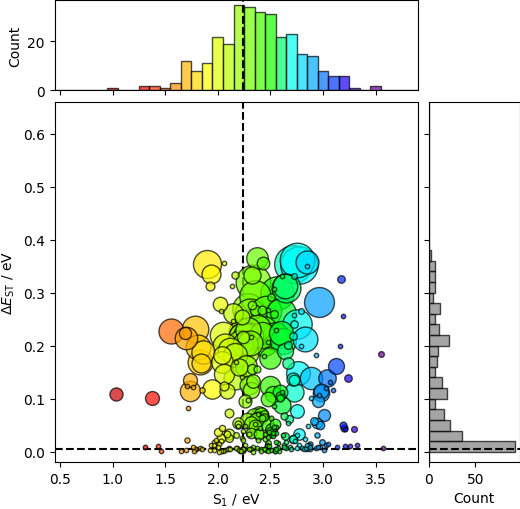}
    \caption{Pm2}
\end{subfigure}
\hspace{2pt}
 \begin{subfigure}[t]{0.48\textwidth}
    \centering
    \includegraphics[scale=0.5]{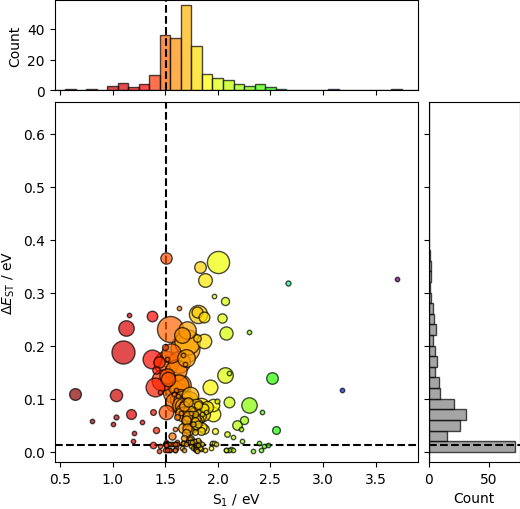}
    \caption{3DMAC-BP-CN}
\end{subfigure}
\hspace{2pt}
\caption{Distribution of S$_1$ and $\Delta E_{\mathrm{ST}}$ values for D-A molecules, where the markers have been coloured according to predicted emission colour and sized according to $f_{\mathrm{S}_1}$. Parent molecule values are indicated with dashed black lines.}
\label{fig:osc_est_mols1}
\end{figure*}

\begin{figure*}[t]
 \centering
 \begin{subfigure}[t]{0.48\textwidth}
    \centering
    \includegraphics[scale=0.5]{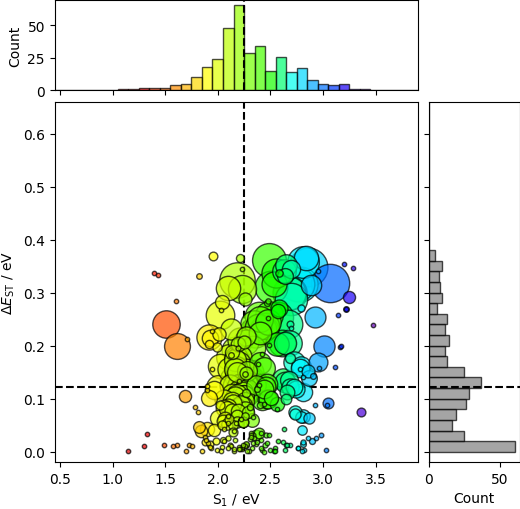}
    \caption{PXZ-TRZ}
\end{subfigure}
\hspace{2pt}
 \begin{subfigure}[t]{0.48\textwidth}
    \centering
    \includegraphics[scale=0.5]{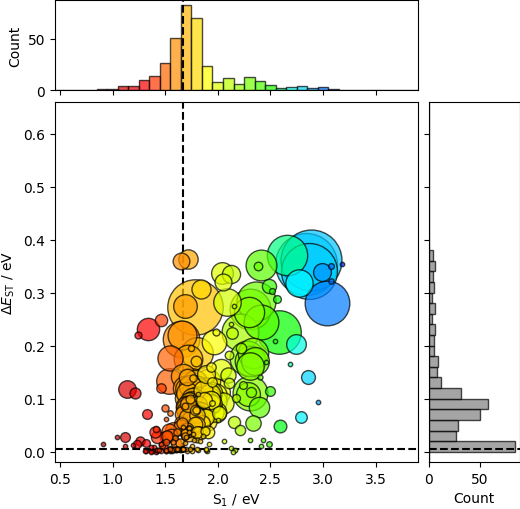}
    \caption{DHPZ-2TRZ}
\end{subfigure}
 \begin{subfigure}[t]{0.48\textwidth}
    \centering
    \includegraphics[scale=0.5]{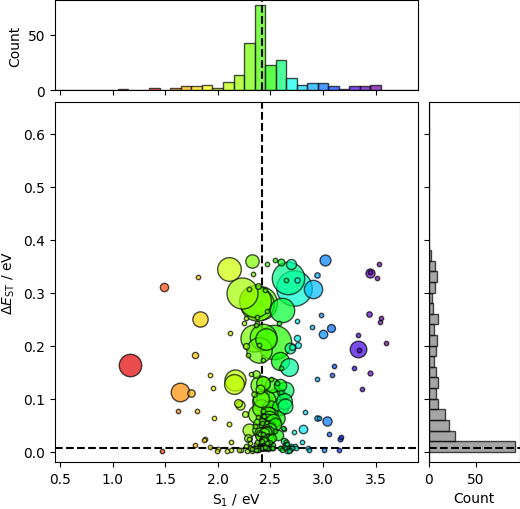}
    \caption{SpiroAC-TRZ}
\end{subfigure}
\hspace{2pt}
 \begin{subfigure}[t]{0.48\textwidth}
    \centering
    \includegraphics[scale=0.5]{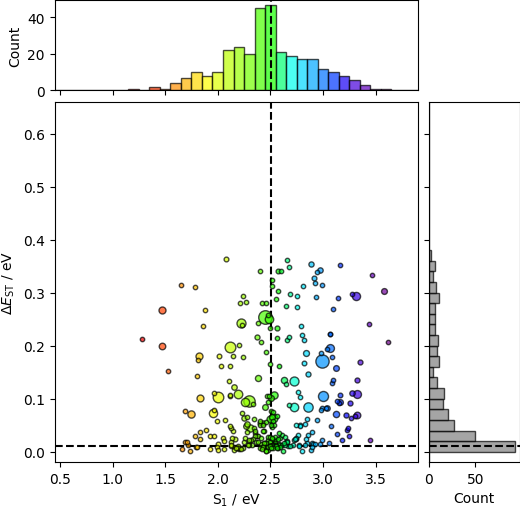}
    \caption{ACRFLCN}
\end{subfigure}
\hspace{2pt}
\caption{Distribution of S$_1$ and $\Delta E_{\mathrm{ST}}$ values for D-A molecules, where the markers have been coloured according to predicted emission colour and sized according to $f_{\mathrm{S}_1}$. Parent molecule values are indicated with dashed black lines.}
\label{fig:osc_est_mols2}
\end{figure*}

\begin{figure*}[t]
 \centering
 \begin{subfigure}[t]{0.48\textwidth}
    \centering
    \includegraphics[scale=0.5]{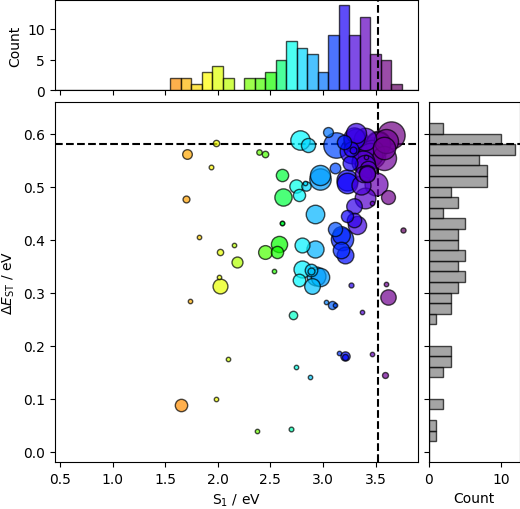}
    \caption{DOBNA}
\end{subfigure}
\hspace{2pt}
 \begin{subfigure}[t]{0.48\textwidth}
    \centering
    \includegraphics[scale=0.5]{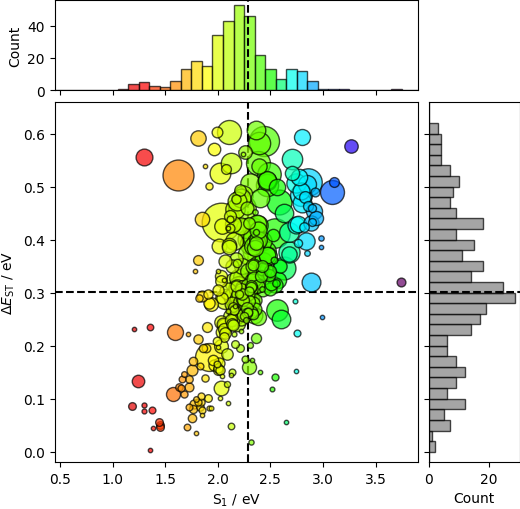}
    \caption{3Cz-DiKTa}
\end{subfigure}
 \begin{subfigure}[t]{0.48\textwidth}
    \centering
    \includegraphics[scale=0.5]{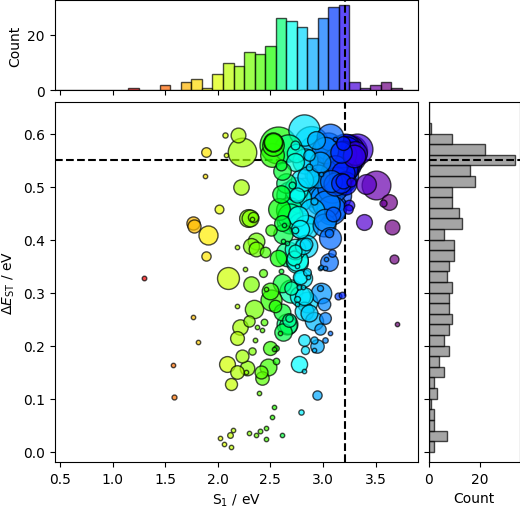}
    \caption{DABNA-1}
\end{subfigure}
\hspace{2pt}
 \begin{subfigure}[t]{0.48\textwidth}
    \centering
    \includegraphics[scale=0.5]{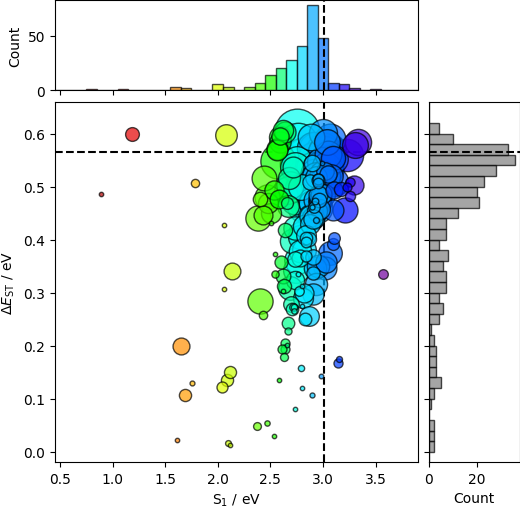}
    \caption{ADBNA-Me-Mes}
\end{subfigure}
 \begin{subfigure}[t]{0.48\textwidth}
    \centering
    \includegraphics[scale=0.5]{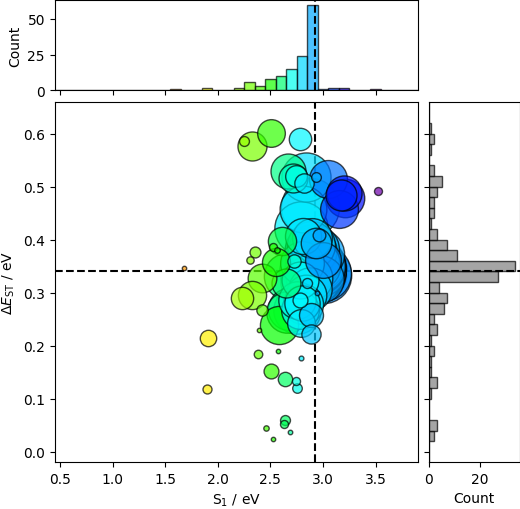}
    \caption{$\nu$-DABNA}
\end{subfigure}
\hspace{2pt}
 \begin{subfigure}[t]{0.48\textwidth}
    \centering
    \includegraphics[scale=0.5]{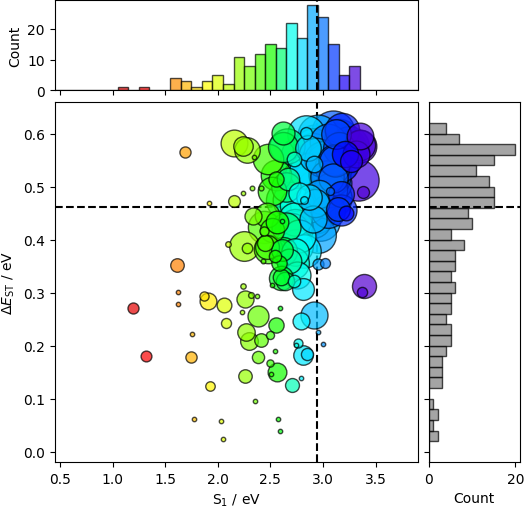}
    \caption{CzBN}
\end{subfigure}
\hspace{2pt}
\caption{Distribution of S$_1$ and $\Delta E_{\mathrm{ST}}$ values for MR molecules, where the markers have been coloured according to predicted emission colour and sized according to $f_{\mathrm{S}_1}$. Parent molecule values are indicated with dashed black lines.}
\label{fig:osc_est_mols3}
\end{figure*}

\begin{figure*}[t]
 \centering
 \begin{subfigure}[t]{0.48\textwidth}
    \centering
    \includegraphics[scale=0.5]{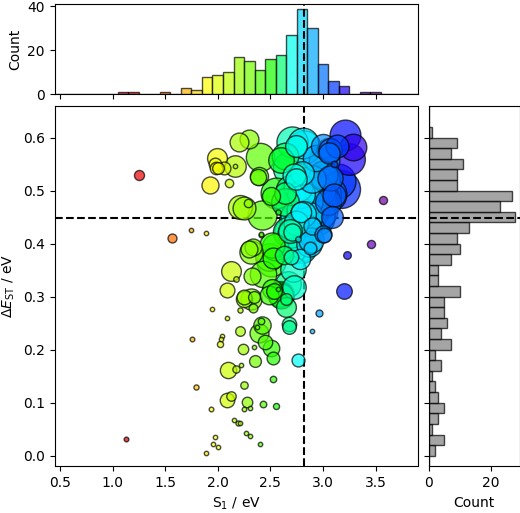}
    \caption{2PTZBN}
\end{subfigure}
\hspace{2pt}
 \begin{subfigure}[t]{0.48\textwidth}
    \centering
    \includegraphics[scale=0.5]{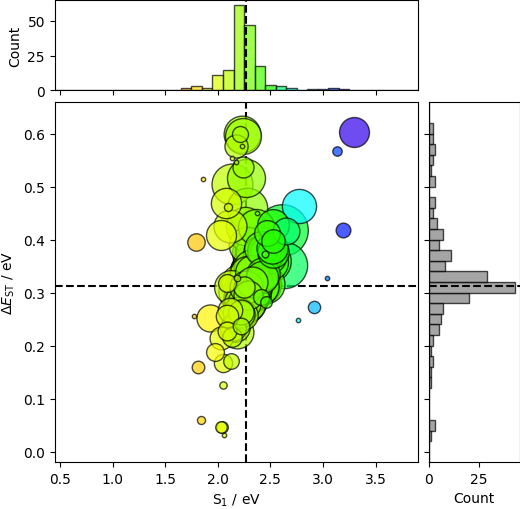}
    \caption{BN3} 
\end{subfigure}
 \begin{subfigure}[t]{0.48\textwidth}
    \centering
    \includegraphics[scale=0.5]{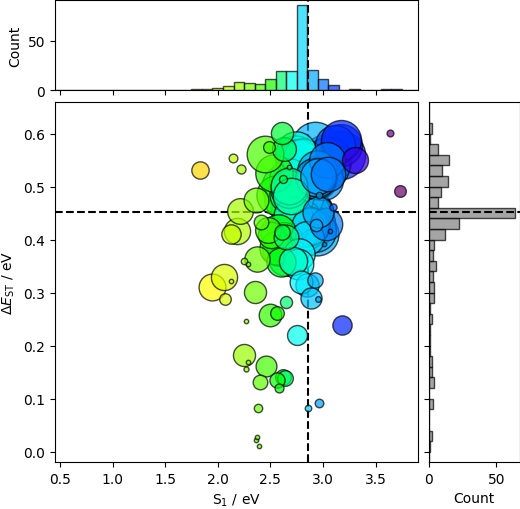}
    \caption{2F-BN}
\end{subfigure}
\hspace{2pt}
 \begin{subfigure}[t]{0.48\textwidth}
    \centering
    \includegraphics[scale=0.5]{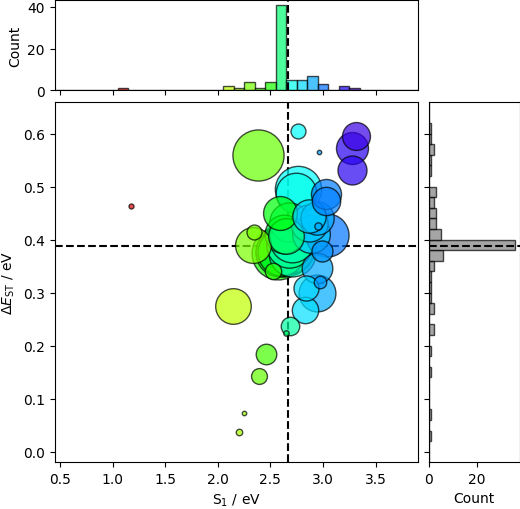}
    \caption{DBNO}
\end{subfigure}
\hspace{2pt}
\caption{Distribution of S$_1$ and $\Delta E_{\mathrm{ST}}$ values for MR molecules, where the markers have been coloured according to predicted emission colour and sized according to $f_{\mathrm{S}_1}$. Parent molecule values are indicated with dashed black lines.}
\label{fig:si_osc_est_mols4}
\end{figure*} 

\clearpage

 \begin{figure}[ht]
    \centering
    \includegraphics[scale=0.5]{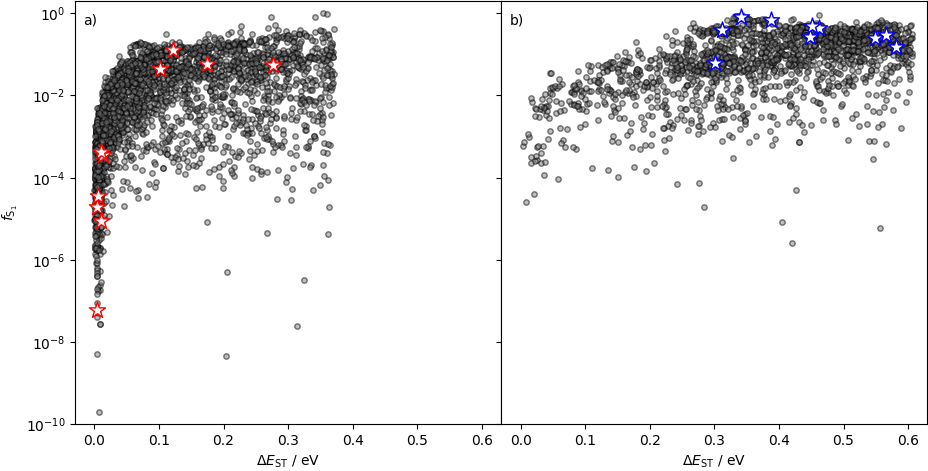}
    \caption{S$_1$ oscillator strength \emph{vs}.\ $\Delta E_{\mathrm{ST}}$ for molecules which passed all workflow filters. Molecules coming from D-A (MR) parent molecules are depicted in a) (b)), while D-A (MR) parent molecules are highlighted with red (blue) $\bigstar$ symbols.
    \label{fig:osc_est}}
\end{figure}

\clearpage

\bibliographystyle{apsrev4-1}
\bibliography{refs}